\documentclass[final,1p,times,authoryear]{elsarticle}
\usepackage{amsmath}
\usepackage{amsthm}
\usepackage{amssymb}
\usepackage{amsfonts}
\usepackage{natbib}
\usepackage[ruled]{algorithm2e}
\setcitestyle{numbers,square}
\newtheorem{theorem}{Theorem}[section] 
\newtheorem{lemma}{Lemma}[section] 
\newtheorem{corollary}{Corollary}[section]

\newtheorem{definition}{Definition}[section] 
 
\newtheorem{example}{Example}

\begin{document}

\begin{frontmatter}

\title{Reachability Analysis of Linear Systems}

\author{Chen shiping}
\address{Sichuan Trade School, Yaan, 625107, Sichuan, China}
\ead{chinshiping@sina.com}
\ead[url]{URL 1}

\author{Ge Xinyu}
\address{Chengdu Institute of Computer Application, Chinese Academy of Sciences, Chengdu, 610000, Sichuan, China}
\address{University of Chinese Academy of Sciences, Beijing, 100049, Beijing, China}
\ead{geeexy@163.com}
\ead[url]{URL 2}

\begin{abstract}
In this paper, we propose a decision procedure of reachability for linear system $\xi'=A\xi+u$, where the matrix $A's$ eigenvalues can be arbitrary algebraic number and the input $u$ is a vector of trigonometric-exponential polynomials. If the initial set contains only one point, the reachability problem under consideration is resorted to the decidability of the sign of trigonometric-exponential polynomial and then achieved by being reduced to verification of a series of univariate polynomial inequalities through Taylor expansions of the related exponential functions and trigonometric functions. If the initial set is open semi-algebraic, we will propose a decision procedure based on openCAD  and an algorithm of real roots isolation derivated from the sign-deciding procedure for the trigonometric-exponential polynomials. The experimental results indicate the efficiency of our approach. Furthermore, the above procedures are complete under the assumption of Schanuel Conjecture.
\end{abstract}
\end{frontmatter}

\section{Introduction}
Hybrid system(HS) combines discrete event systems with differential equations in a manner that is ideal for the modeling, analysis, and design of embedded systems. To guarantee the correctness of these systems is vital so that we can bet our lives on them (see [26]). The safety criticality of many applications requires the use of formal methods to ensure that an unsafe region of the state space is not reachable from a set of initial conditions. This makes the reachability problem for hybrid systems very important. It should be pointed out that the reachability problem of most of HSs is undecidable (see [14]) except for some simple cases.

[15] investigated vector fields of the following linear system
\begin{equation}
\xi'=A\xi+u
\end{equation}
where $\xi(t)\in{\mathbb{R}^{n}}$ is the state of the system at time $t$,\ $A\in \mathbb{R}^{n\times n}$ is the system matrix, and $u:\mathbb{R} \to \mathbb{R}^n$ is a piecewise continuous function which is called the input.

Given an initial state $\xi(0)=x=(x_1,x_2,\cdots,x_n)$, the solution to (1) at time $t\geq 0$ is denoted by $\operatorname{\xi}(t)=\phi (x,t)$. Then the forward reachable set $\operatorname{Post}(X)$ of (1) from a given initial set $X$ is defined as:
\begin{equation}
\operatorname{Post}(X)=\{y\in \mathbb{R}^{n}\mid \exists x \exists t:x \in X\land t\geq 0\land \phi (x,t)=y\}.
\end{equation}

Now, the problem under consideration is formulated as follows: Given a linear system, an initial set \ $X$ and an unsafe set \ $Y$,  the problem is to verify whether any unsafe state in \ $Y$ is not reachable by any trajectory starting from \ $X$, i.e., whether $\operatorname{Post}(X)\cap Y= \emptyset $, or $\operatorname{F}(X,Y)={\exists x\exists y\exists t:x \in X\land y\in Y\land t\geq 0\land \phi (x,t)=y}$. If the unsafe set \ $Y$ is semi-algebraic, the problem can be transformed to decide whether all the inequalities in $Y$ hold.

A set $X \subset \mathbb{R}^{n}$ is said semi-algebraic if it is defined as $\{x\in \mathbb{R}^{n}\mid p_{1}(x)\Delta 0,\cdots, p_{n}(x) \Delta 0\}$, for some polynomials $p_{1}(x),\cdots,p_{n}(x)\in \mathbb{R}[x]$, where $\Delta\in \{\geq ,>\}$. A semi-algebraic set $X$ is said to be open if all $\Delta$s are instantiated to $>$.

[15] obtained the decidability of the reachability problems for the following three families of vector fields:

\begin{enumerate}
\item $A$ is nilpotent, i.e. $A^{n} = 0$, and each component of $u$ is a polynomial;

\item $A$ is diagonalizable with rational eigenvalues, and each component of $u$ is of the form $\Sigma_{i=1}^{m}c_{i} e^{\lambda_{i}t}$, where $\lambda_{i}$s are rational and $c_i$s are subject to semi-algebraic constraints;

\item $A$ is diagonalizable with purely imaginary rational eigenvalues, and each component of u is of the form $\Sigma_{i=1}^{m} (c_{i} \sin(\lambda_{i} t)+d_{i} \cos(\lambda_{i} t))$, where $\lambda_i$s are rationals and $c_{i}$s and $d_{i}$s are subject to semi-algebraic constraints.
\end{enumerate}

[7] and [12] generalize the case 2 above by different schemes independently as following:

\ $A$ is diagonalizable with real eigenvalues, and each component of $u$ is of the form $\Sigma_{i=1}^{m}c_{i}e^{\lambda_{i}t}$, where $\lambda_{i}$s are reals and $c_i$s are subject to semi-algebraic constraints.

For \ $\lambda_{i}$s are real algebraic numbers (Denoted as \ $\mathbb{R}_{alg}$, the same below), and under the assumption of Schanuel’s Conjecture, their algorithms are complete. To the best of our knowledge, these results are the strongest ones on the decidability of the reachability problems of HSs obtained so far.

 In this paper, we will generalize the linear system by follows.

The matrix $A'$s eigenvalues can be arbitrary algebraic numbers, and the the input u is a vector of real trigonometric-exponential polynomials, the initial set $X$ contains only one point or is semi-algebraic, the unsafe set $Y$ is semi-algebraic. Meanwhile, we care the reachability of system within a specific bounded interval only.

A term with the following form is called a trigonometric-exponential polynomial(TEP), $\Sigma_{i=1}^s e^{u_i x} (f_i (x) \sin(v_i x)+g_i (x) \cos(r_i x))$, where $f_i(x),g_i(x) \in{\mathbb{C}[x]}$,\ $u_i$, $v_i$, $r_i \in \mathbb{A}$. If a TEP is real-valued, it is called real trigonometric-exponential polynomial(RTEP).

For initial point \ $\xi(0)= x_0=(x_1,\cdots,x_n)$, the solution of system (1) is \ $\xi_k(t)=\Phi_k(x_0,t)=\Sigma_i^s e^{u_i t}(f_i(x_0,t) \sin(v_i t)+g_i(x_0,t) \cos(r_i t))$, where \ $f_i(x_0,t)$, $g_i(x_0,t)$ are polynomials of $t$ with parameter  \ $(x_1,\cdots, x_n)$,\ $u_i$, $v_i$, $r_i$ are real algebraic numbers. That is to say, \ $\xi_k(t)$ is a RTEP with parameter \ $x_0=(x_1,\cdots, x_n)$.

As each polynomial of the semi-algebraic unsafe set \ $q_{k}(y_1,\cdots ,y_n)\in \mathbb{R}[y_1,\cdots, y_n]$, so \ $q_{k}(\phi_{1}(x_0,t), \cdots, \phi_{n}(x_0,t))$ is a RTEP with parameter $x_0=(x_1,\cdots,x_n)$. So, to obtain the decidability, we can resort to the sign decidability of the transcendental function of a class of trigonometric-exponential polynomials with parameters.

In recent years, the sign deciding of transcendental function, or the automatic proof of transcendental inequalities is one of the key points and difficulties pursued by scholars.

Lately, a so-called Taylor-substitution method is employed to deal with problems of transcendental functions. [4,6] solved the problem of automated proof of mixed trigonometric-polynomial inequalities defined by the formula $f(x)=\Sigma a_{i}x^{p_i} \sin^{q_{i}}(x) \cos^{r_{i}}(x) \\ > 0$ and the exponential polynomial inequalities of form $f(x,e^{-x})>0$, reducing the proof of the original inequality to a series of verification of univariate polynomial inequality by Taylor expansion of the inverse tangent function $\arctan(x)$ or exponential function \ $e^{-x}$, without discussing the existence of real roots of transcendental functions directly. Of course, isolating the real zeros of mixed trigonometric-polynomial and exponential polynomial can be easily implemented based on the above method.

A general transcendental function may contain more than one transcendental factors, and the function after Taylor-substitution once may still contain transcendental factors, so Taylor-substitution need to be performed again or even many times. [5] proposed a procedure named Successive Taylor-substitution  to solve the automatic proof of a class of generalized polynomial  of the form \ $F(x)= f(x,trans_{1}(x),\cdots, trans_{t}(x))>0$, containing more than one transcendental factors.

The essence of the sign-deciding of a given real function is the existence and classification of real zeros of the function. There are some wonderful achievements for isolating real roots of certain classes of transcendental functions.

[17] presented a decision procedure for a certain class of sentences of first order logic involving integral polynomials and a certain specific analytic transcendental function \ $trans(x)$ in which the variables range over the real numbers (See also [1]). The list of transcendental functions to which the decision method directly applies includes \ $e^{x}$, the exponential function with respect to base \ $e$, \ $\ln(x)$, the natural logarithm of $x$, and \ $\arctan(x)$, the inverse tangent function. In the case \ $trans(x)= e^x$, the decision procedure has been implemented in the computer logic system REDLOG. The decision method is based upon an algorithm for isolating the real zeros of a certain kind of generalised integral polynomial in \ $trans(x)$, $f(x,trans(x))$, where \ $f(x,y)$ is a given polynomial in \ $y$ whose coefficients are elements of the ring of fractions of \ $\mathbb{Z}[x]$ with respect to powers of a specific integral polynomial \ $d(x)$, which uses pseudo-differentiation and Rolle's Theorem, and also relies upon a classical result of Lindemann Theorem.

Based on a generalized Budain-Fourier Theorem, [4] presented a real root isolation procedure, which is easily adapted to decide the sign of the corresponding function, for more expressive exp-log-arctan functions obtained by composition and rational operations from \ $exp$, \ $ln$, \ $\arctan$ and real constants. By Strzebonski's result, we can conclude that there exists algorithm to decide the sign of trigonometric-exponential polynomial on a given interval. But the procedure employed an algorithm of [10] to determine signs of exp-log-arctan functions at simple roots of other exp-log-arctan functions, and the proof that the algorithm of [10] terminates relies on Schanuel’s Conjecture. It is also pointed out that it is possible to create problems involving very large or very small numbers which will require such large precision that their solution will be infeasible. Furthermore, the implementation of [24]'s procedure does not use the zero testing algorithm defined in [10], a zero testing heuristic is used instead.

[11] proposes an algorithm to isolate all real roots for a exponential polynomial of the form $f(t) = \Sigma_{i=0}^s f_i(t) e^{v_i t}$ based on Differential-mean-value Theorem (i.e., Rolle's Theorem), where $s\in\mathbb{N}$, \ $f_i(t)\in \mathbb{R}[t]$ and $v_i \in\mathbb{R}$ are pairwise different. The algorithm's completeness depends on Schanuel’s Conjecture.

[7] consider a class of univariate real functions, poly-powers, with the normal form $f(x) = \beta_0 + \beta_1 x^{\alpha_1} +\cdots + \beta_n x^{\alpha_n}$, where $\alpha_i (1 \leq i \leq n)$ and $\beta_i (0 \leq i \leq n)$ are real algebraic numbers with $0 < \alpha_1 < \cdots < \alpha_n$ and $n > 0$. They perform factorization to simplify the input poly-power and classify poly-powers into simple and non-simple ones, depending on the number of linearly independent exponents in an irreducible poly-power. For the former, they present two complete isolation algorithms, exclusion and differentiation. For the latter, they repeatedly differentiate poly-powers until the resulting poly-power has at most two terms, and lifts isolation intervals of derivatives to those of the original poly-power. The whole procedures are established on algebraic manipulation, and hence are absolutely exact. The completeness is ensured by Gelfond-Schneider Theorem for simple poly-powers, and by Schanuel Conjecture for non-simple ones.

[8] presents a solution to the continuous Skolem Problem whether a real-valued function satisfying a linear differential equation, $f^{(n)}+a_{n-1}f^{(n- 1)}+\cdots +a_{0}f =0$, has a zero in a given interval. For general cases, the characteristic roots of the linear differential equation have the form \ $\lambda_j = r_j + I \omega_j $, where $r_j$, $\omega_j \in \mathbb{R}$, and  \ $I^2=-1$, then \ $f(t)$ can be  wrote  in the form \ $f(t) = \sum_{j=1}^m e^{r_{j} t} (Q_{1,j} (t) \sin(\omega_j t) + Q_{2,j} (t) \cos(\omega_j t))$, where the polynomials \ $Q_{1,j}$, \ $Q_{2,j}$ have real-algebraic coefficients. Chonev's procedure is based on the analysis of the exponential polynomial \ $f(t)=P(t,e^{a_{1}t},\cdots ,e^{a_{r}t},e^{I b_{1}t},\cdots,e^{I b_{s}t})$ and the M-Lipschitz condition of real function. The completeness also relies on Schanuel Conjecture.

In this paper, we will follow the idea of [4-6] and combine the methods of the above papers. A so-called Successive Taylor-substitution procedure to decide the sign of the trigonometric-exponential polynomials will be proposed by performing Taylor-Substitution repeatedly to reduce the problem to a series of verification of  polynomial inequalities with only one variable. Similar to [24], [7] and [11], the algorithm is complete under Schanuel Conjecture. And then the sign-deciding algorithm will be employed to isolation of real roots of trigonometric-exponential polynomial and the reachability analysis of linear systems. If the initial set of linear system contains only one point, the sign-deciding algorithm can decide the reachability of linear system directly. If the initial set of linear system is an open semi-algebraic set, we give a decision procedure with the help of openCAD (see [13]) and an algorithm of real root isolation of trigonometric-exponential polynomial derivated from the  sign-deciding algorithm.

The rest of the paper is organized as follows. Section 2 studies the properties of exponential polynomial under the assumption of Schanuel Conjecture and presents a procedure to factorize the trigonometric-exponential polynomials without multiple roots. Section 3 proposes the Successive Taylor-substitution algorithm for deciding the sign of transcendental function polynomial and discuss its completeness. Section 4 presents deciding procedure of reachability analysis of linear systems on two situations depending on whether the initial set of linear system contains only one point or the initial set is an open semi-algebraic set. We will conclude the paper in Section 5.

\section{Exponential polynomials}

\subsection{Exponential polynomials with simple roots}
For an \ $(n+1)$-ary polynomial ring \ $\mathbb{K}[x, x_{1},\cdots, x_{n}]$  on a given field \ $\mathbb{K}$, define a mapping \ $hom:f(x, x_{1},\cdots, x_{n})\rightarrow F(x) = f(x, trans_{1}(x),\cdots, trans_{n}(x))$, substituting \ $x_{i}$ with a transcendental function \ $trans_{i}(x)$ for $i=1,\cdots, n$.

\begin{definition}
For an $(n+1)$-ary polynomial $f(x, x_{1},\cdots, x_{n})$ in a given filed $\mathbb{K}$, its image under mapping $hom$ $F(x)=hom(f)=f(x, trans_{1}(x),\cdots, trans_{n}(x))$ is called a transcendental polynomial, and $trans_{i}(x)$ is called transcendental factor.
\end{definition}

For example,\ $x+\sin(x)$,\ $x+\cos(\sqrt{x})$,\ $x+e^{\sin(x)}$,\ $x+\arctan(x)-\arcsin(x)$ are transcendental polynomials, $\sin(x)$, $\cos(\sqrt{x})$, $e^{\sin(x)}$, $\arctan(x)$, $\arcsin(x)$ are their transcendental factors.

For \ $F(x)=hom(f)=f(x, trans_{1}(x),\cdots, trans_{n}(x))$,\ $f\in{\mathbb{C}[x, y_{1},\cdots,y_{n}]}$, if all $trans_{i}(x)$s are exponential like $e^{v_{i}x}$, where \ $v_{i}\in\mathbb{A}$,\ $F(x)$ is called an exponential polynomial (short for EP), on this occasion, the mapping $hom$ is also denoted more accurately by \ $hom[v_1,\cdots,v_n]$, i.e. \ $hom[v_1,\cdots,v_n](f(x,x_1,\cdots,x_n))=f(x,e^{v_1},\cdots,e^{v_n})$.

For \ $F(x)=hom(f)=f(x, trans_{1}(x), \cdots , trans_{n}(x))$, \ $f\in{\mathbb{C}[x, y_{1},\cdots,y_{n}]}$, if all \ $trans_{i}(x)$s are exponential like \ $e^{u_{i}x}$ , or sine like \ $\sin(v_{i}x)$, or cosine like \ $\cos(r_{i}x)$, where $u_{i},v_{i},r_{i}\in \mathbb{R}_{alg}$, \ $F(x)$ is called a trigonometric-exponential polynomial (short for TEP). For general, a TEP has the formal form \ $f(x,e^{u_{1}x},\cdots,e^{u_{r}x},\sin(v_{1}x),\cdots, \sin(v_{s}x),\cos(r_{1}x),\cdots,\cos(r_{t}x))$, where \ $f(x,y_1, \cdots, y_{r}, z_{1}, \cdots, z_{s}, w_{1}, \cdots, w_{t})$ is a \ $(r+s+t+1)$-ary polynomial in field \ $\mathbb{K}$. Of course, this definition for trigonometric-exponential polynomial is consistent with that in the previous section.

\begin{definition}
The complex numbers $a_{1}, \cdots , a_{n}$ are called linearly independent over $\mathbb{Q}$, if for all \ $c_{1}, \cdots, c_{k} \in \mathbb{Q}$ such that $\sum{}_{i=1}^{k}c_{i} a_{i}=0$, we have \ $c_{1}= \cdots = c_{k} = 0$.
\end{definition}

\begin{definition}
The complex numbers set \ $A=\{a_{1}, \cdots , a_{n}\}$ is an integral basis of  complex numbers set \ $B=\{b_{1}, \cdots , b_{m}\}$, if \ $A$ is  linearly independent over $\mathbb{Q}$ and for each \ $i$, \ $1\leq i \leq m$ , there exits \ $p_1, \cdots, p_n \in \mathbb{Z}$ such that \ $b_i = \sum_{j=1}^n p_j a_j$.  \qed
\end{definition}

\begin{definition}
$\{v_1,v_2,\cdots,v_n\}$ is called a basis of exponential polynomial \ $F(x)$, if there exists a polynomial \ $f(x,x_1,\cdots,x_n)\in \mathbb{K}[x,x_1,\cdots,x_n]$ such that \ $F(x)=hom[v_1,\cdots,v_n](f)$. If \ $v_{1}, v_{2},\cdots, v_{n} \in \mathbb{A}$ is linearly independent over $\mathbb{Q}$, $\{v_{1},v_{2}, \cdots, v_{n}\}$ is called a regular basis of \ $F(x)$.
\end{definition}

Obviously, if $B$ is a basis of an exponential polynomial $F(x)$ and \ $A$ is an integral basis of \ $B$, then \ $A$ is a regular basis of \ $F(x)$.

An exponential polynomial $F(x) = f(x, e^{v_{1} x}, \cdots, e^{v_{n} x})$ can also be expressed as $F(x)=\sum_{0}^{m}f_{i}(x) e^{\lambda_i x}$, where $f_{i}(x)\in{\mathbb{K}[x]}$. Furthermore, there may be different expressions for the same exponential polynomial. For example, $F(x) = x+e^{(\sqrt{2}+1)x} = hom[\sqrt{2}+1](f_{1}) = hom[\sqrt{2}, 1](f_{2})$, where $f_1 = x + y_1$, $f_{2} = x+y_{1}y_{2}$.

\begin{definition}
For $F(x)=\sum_{0}^{m}f_{i}(x)e^{\lambda_i x}$, $gcd(f_{0}(x), f_{1}(x),\cdots, f_{n}(x))$ is called the content of \ $F(x)$, denoted by $\operatorname{cont}(F)$.
\end{definition}

\begin{definition}
The complex numbers $a_{1}, \cdots, a_{n}$ are algebraically independent if there is no non-zero $n$-ary polynomial $f\in\mathbb{A}[x_{1}, \cdots, x_{n}]$ such that $f(a_{1},\cdots, a_{n}) = 0$.
\end{definition}

\begin{lemma}
If $f(x,y_{1},\cdots, y_{n})\in{\mathbb{A}[x,y_{1}, \cdots, y_{n}]}$, $F(x)=f(x, e^{v_{1} x},\cdots, e^{v_{n} x})$ and \ $\operatorname{cont}(F(x))=1$, \ $v_{1},\cdots,v_{n}\in{A}$ are linearly independent over $\mathbb{Q}$, then for each rational $x_{0}\not= 0$ and \ $ x_{0}\in\mathbb{Q}$, $F(x_{0})\not= 0$.
\begin{proof}
Let $g(y_{1},\cdots,y_{n})=f(x_{0},y_{1},\cdots, y_{n})$, then $g\in\mathbb{A}[y_{1},\cdots,y_{n}]$ and is not identically zero due to the assumption that \ $\operatorname{cont}(F)=1$. Meanwhile, $v_{1},\cdots, v_{n}\in\mathbb{A}$ are linearly independent over $\mathbb{Q}$, so $v_{1}x_{0},\cdots, v_{n}x_{0}\in\mathbb{A}$ are also linear independent over $\mathbb{Q}$. By Lindemayn-Weierstrass Theorem, $e^{v_{1}x_{0}},\cdots,e^{v_{n}x_{0}}$ are algebraical independent, i.e. $g(e^{v_{1}x_{0}},\cdots, e^{v_{n}x_{0}})=F(x_{0})\not=0$.
\end{proof}
\end{lemma}

\begin{lemma}
If $f(x, y_{1},\cdots, y_{n})\in \mathbb{A}[x, y_{1},\cdots, y_{n}]$ is not identically zero, $v_{1},\cdots, v_{n}\in\mathbb{A}$ are linear independent over $\mathbb{Q}$, Then $F(x)=f(x,e^{v_{1}x}, \cdots, e^{v_{n}x})$ is not identically zero.
\begin{proof}
Let \ $\operatorname{cont}(F(x))=f_{1}(x)$, that is to say,\ $f(x,y_{1}, \cdots, y_{n})=f_{1}(x)\times f_{2}(x,y_{1}, \cdots, y_{n})$, $f_{1}(x)$ is polynomial of \ $x$ and $\operatorname{cont}(f_{2}(x,e^{v_{1}x}, \cdots,e^{v_{n}x}))=1$.

Let \ $x_{0}\in\mathbb{A}$ such that \ $0\not=x_{0}\in\mathbb{Q}$ and \ $f_{1}(x_{0})\not=0$. By Lemma 2.1 \ $f_{2}(x_{0}, e^{v_{1}x_{0}},\cdots ,e^{v_{n}x_{0}})\not=0$. So \ $F(x_{0})\not=0$.
\end{proof}
\end{lemma}

Denote $\operatorname{EXP}[v_{1}, v_{2},\cdots,v_{n}]=\{F(x)=hom[v_{1}, v_{2},\cdots,v_{n}](f),f\in\mathbb{A}[x, y_{1},\cdots,y_{n}]\}$.

\begin{theorem}
If $v_{1},\cdots,v_{n} \in\mathbb{A}$ are linearly independent over 
$\mathbb{Q}$, the mapping $hom[v_{1}, v_{2},\cdots, v_{n}]$ 
is bijective from  $\mathbb{A}[x, y_{1},\cdots,y_{n}]$ to $\operatorname{EXP}[v_{1}, v_{2},\cdots, v_{n}]$, 
so it's inverse mapping $hom[v_{1},v_{2},\cdots, v_{n}]^{-1}$ exists.
\end{theorem}

For the same exponential polynomial $F(x)$,\ $hom[B_{1}]^{-1}(F)$ and\ $hom[B_{2}]^{-1}(F)$ may be different if \ $B_{1}\not= B_{2}$. For example, Let \ $F(x)=x+e^{\sqrt{2}x+x}$, \ $hom[\sqrt{2}+1]^{-1}(F(x))=x+y_{1}$,\  $hom[\sqrt{2},1]^{-1}(F(x))=x+y_{1}y_{2}$.

$\textbf{Schanuel Conjecture}$ \ If \ $x_1,\cdots, x_n \in \mathbb{C}$ are linearly independent over \ $\mathbb{Q}$, then there are at least \ $n$ algebraically independent numbers among $x_1,\cdots, x_n, e^{x_1},\cdots, e^{x_n}$.

The Schanuel Conjecture was proposed by S.H.Schanuel in \ $1960$, which is is generally considered to be correct but hasn't been proved.
The following discussions are under the assumption of Schanuel Conjecture.

\begin{lemma}
Suppose \ $v_1,\cdots, v_n \in \mathbb{A}$ are linearly independent over $\mathbb{Q}$ and \ $t\not=0$, then there are at least \ $n$ algebraically independent numbers among \ $t,e^{v_{1}t},\cdots, e^{v_{n}t}$.
\begin{proof}
\ $v_1,\cdots, v_n$ are linearly independent over \ $\mathbb{Q}$, so are \ $v_{1}t,\cdots, v_{n}t$. By Schanuel Conjecture, there are at least \ $n$ algebraically independent numbers among \ $v_{1}t,\cdots,v_{n}t,e^{v_{1}t} ,\cdots ,e^{v_{n}t}$ . As \ $v_1,\cdots ,v_n\in\mathbb{A}$, there are at least \ $n$ algebraically independent numbers among \ $t,e^{v_{1}t} ,\cdots ,e^{v_{n}t}$.
\end{proof}
\end{lemma}

\begin{lemma}
If $f_1(x, y_1,\cdots, y_n)$, $f_2(x, y_1,\cdots , y_n)\in \mathbb{A}[x, y_1,\cdots, y_n]$ are co-prime, suppose \ $v_1, \cdots,  \\  v_n \in \mathbb{A}$ are linear independent over $\mathbb{Q}$, then \ $F_1(x)=f_1(x, e^{v_{1}x},\cdots , e^{v_{n}x})$ and $F_2(x)=f_2(x, e^{v_{1}x},\cdots, \\ e^{v_{n}x})$ have no common roots other than $0$.
\begin{proof}
Suppose \ $F_1(x)$ and \ $F_2(x)$ have common root \ $x_0$ and \ $x_0\not=0$. By Lemma 2.3, there are at least \ $n$ algebraically independent numbers among \ $x_0 ,e^{v_{1}x_{0}} ,\cdots ,e^{v_{n}x_{0}}$. Without losing generality, suppose \ $\{x_0 ,e^{v_{1}x_{0}},\cdots ,e^{v_{n-1}x_{0}}\}$ is algebraically independent.

$f_1(x_0,e^{v_{1}x_{0}},\cdots, e^{v_{n-1}x_{0}},y_n)$ and \ $f_2(x_0, e^{v_{1}x_{0}},\cdots, e^{v_{n-1}x_{0}},y_n)$ are two polynomials of \ $y_n$, which have common root \ $e^{v_{n}t_{0}}$. Let \ $g(x,y_1,\cdots,y_{n-1})=\operatorname{resultant}(f_1(x,y_1,\cdots, y_n), f_2(x,y_1,\cdots, y_n) ,y_n)$, then $g\in \mathbb{A}[x, y_1,\cdots, y_{n-1}]$ and $g(x_0 ,e^{v_{1}x_{0}} ,\cdots ,e^{v_{n-1}x_{0}} ) =0$. As \ $f_1(x, y_1,\cdots, y_n)$ and \ $f_2(x, y_1,\cdots, \\ y_n)$ are co-prime, \ $g(x,y_1,\cdots, y_{n-1})$ can not be identically zero. But $g(x_0 ,e^{v_{1}x_{0}},\cdots, e^{v_{n-1}x_{0}} ) =0$ contradicts with the fact that \ $\{x_0 ,e^{v_{1}x_{0}} ,\cdots ,e^{v_{n-1}x_{0}}\}$ are algebraically independent. We conclude that Lemma 2.4 holds.
\end{proof}
\end{lemma}

\begin{theorem}
If $f(x, y_{1},\cdots, y_{n})\in \mathbb{A}[x, y_{1},\cdots, y_{n}]$ is irreducible, suppose $v_{1},\cdots, v_{n}\in A$ are linear independent over \ $\mathbb{Q}$, then \ $F(x)= f(x, e^{v_{1}t},\cdots, e^{v_{n}t})$ has no multiple roots other than \ $0$.
\begin{proof}
Denote \ $f'(x,y_{1},\cdots , y_{n}) = hom[v_{1},v_{2},\cdots ,v_{n}]^{-1}(F'(x))$, then \ $f'(x,y_{1},\cdots, y_{n})=f'_{x}+\sum v_{i}y_{i}f'_{y_{i}}$, so $degree(f',x)\leq degree(f,x)$,\ $degree(f',y_{i}) \leq  degree(f,y_{i})$ for \ $i=1,\cdots ,n$. As \ $f(x,y_{1}, \cdots , y_{n})$ is irreducible, so \ $f(x,y_{1},\cdots , y_{n})$ and \ $f'(x,y_{1},\cdots , y_{n})$ are co-prime. Then \ $F(x)$ and \ $F'(x)$ have no common roots other than \ $0$. We conclude that Theorem 2.2 holds.
\end{proof}
\end{theorem}

\begin{corollary}
If \ $f(x, y_{1},\cdots, y_{n})\in \mathbb{A}[x, y_{1},\cdots, y_{n}]$ is square-free, suppose \ $v_{1},\cdots, v_{n}\in \mathbb{A}$ are linear independent over \ $\mathbb{Q}$, then \ $F(x)= f(x, e^{v_{1}x}, \cdots, e^{v_{n}x})$ has no multiple roots other than \ $0$.
\end{corollary}

\begin{definition}
For $F(x)=f(t, e^{v_{1}t},\cdots, e^{v_{n}t})$, if there exists one regular basis $\lambda_{1}$,$\cdots$, $\lambda_{r}$ such that \ $hom[\lambda_{1},\cdots, \lambda_{r}]^{-1}(F(x))$ is square-free, then \ $F(x)$ is called square-free.
\end{definition}

\begin{corollary}
If \ $F(t)= f(t, e^{v_{1}t},\cdots, e^{v_{n}t})$ is square-free, then \ $F(t)$ has no multiple roots other than \ $0$.
\end{corollary}

\subsection{The TEP's factorization without multiple roots}

As the algorithms of the next two sections will work well if all roots are simple, we consider the problem of reducing root multiplicities for TEP in this section. It will be shown that each TEP can be replaced by another TEP with the same zeros  such that all zeros are simple.

By Euler Formula,\ $\sin(x)=\frac{e^{I x}-e^{-I x}}{2I}$, \ $\cos(x)=\frac{e^{I x}+e^{-I x}}{2}$, a TEP can be described in the form of \ $f(x, e^{u_{1}x},\cdots, e^{u_{r}x}, e^{I v_{1}x}, e^{-I v_{1}x},\cdots, e^{I v_{s}x},e^{-I v_{s}x})$, where \ $f\in\mathbb{C}[x, y_1, \cdots , y_r, z_{11}, z_{12}, \cdots , z_{s1}, z_{s2}]$ and \ $u_{1}, \cdots, u_{r}, v_1, \cdots, v_{s}\in \mathbb{R}_{alg}$, \ $I^2=-1$.

Let \ $y_i=e^{u_{i}x}$,\ $z_i=e^{I v_{i} x}$, then a TEP can be transformed to a Laurent polynomial with the form as \ $f(x, y_1, \cdots , y_r, z_1, z_{1}^{-1}, \cdots , z_s, z_{s}^{-1})$. Denotes \ $\mathbb{LR} := \mathbb{C}[x, y_1, \cdots , y_r, z_1, z_{1}^{-1}, \cdots , z_s, z_{s}^{-1}]$, thus for any \ $P\in \mathbb{LR}$, there exists \ $Q\in \mathbb{C}[x, y_1,\cdots , y_r, z_1,\cdots , z_s]$ such that \ $P=Q/Z^{\vec{p}}$, \ $Q$ and \ $Z^{\vec{p}}$ are co-prime, where \ $\vec{p} = <p_1,\cdots, p_s>$ is a tuple of integers and \ $Z^{\vec{p}}= z_{1}^{p_1}\cdots z_{s}^{p_s}$, the same below. If \ $factor(Q)$ is a factorization of the polynomial \ $Q$, we call \ $factor(Q)/Z^{\vec{p}}$ or \ $Z^{-\vec{p}}factor(Q)$ a factorization of \ $P\in \mathbb{LR}$, where \ $-\vec{p} = <-p_1,\cdots, -p_s>$.

For \ $P(x, y_1,\cdots, y_r, z_1,\cdots, z_s) = \sum_{j=1}^{n}a_j(x) y_{1}^{v_{j1}}\cdots y_{r}^{v_{jr}} z_{1}^{w_{j1}}\cdots z_s^{w_{js}}$, if \ $a_j(x)\in \mathbb{C}[x]$, \ $v_{j1},\cdots, v_{jr}\in \mathbb{Z}$,\ $w_{j1},\cdots, w_{js}\in \mathbb{Q}$ for \ $j=1,\dots,n$, \ $P$ is called a generalized Laurent polynomial(\ $\mathbb{GLR}$). For \ $P\in \mathbb{GLR}$, \ $\{u_1, \cdots , u_r\}$ and \ $\{v_1, \cdots , v_s\}$ are real algebraic, denote
\begin{equation}
\begin{aligned}
& \operatorname{LRhom}[u_1, \cdots , u_r; v_1,\cdots , v_s](P) = P(x, e^{u_{1} x},\cdots, e^{u_{r} x}, e^{Iv_{1}}, \cdots, e^{Iv_{s}}) \\
&= \sum_{j=1}^{n}a_j(x) e^{v_{j1} u_{1} x}\cdots e^{v_{jr} u_{r} x} e^{I w_{j1} v_{1} x}\cdots e^{I w_{js} v_{s} x} \\
&= \sum_{j=1}^{n}a_j(x) e^{v_{j1} u_{1} x}\cdots e^{v_{jr} u_{r} x} (\cos(w_{j1} v_{1}x)+I\sin(w_{j1} v_{1}x))\cdots (\cos( w_{js} v_{s}x)+I\sin(w_{js} v_{s}x)),
  \nonumber
\end{aligned}
\end{equation}
in the case of not causing misunderstanding, abbreviated as\ $\operatorname{LRhom}(P)$. Obviously, $\operatorname{LRhom}(P)$ is a TEP.

For example, if \ $P=x+y_1+z_1+z_2^{-1}$, then \ $\operatorname{LRhom}[\sqrt{2}; \sqrt{3}, \sqrt{5}](P) =x+e^{\sqrt{2}x}+ e^{I\sqrt{3}x} + e^{-I\sqrt{5}x}=x+e^{\sqrt{2}x}+\cos(\sqrt{3}x)+ I\sin(\sqrt{3}x) + \cos(\sqrt{5}x)-I\sin(\sqrt{5}x)$.

\begin{theorem}\label{theorem2.3}
  If \ $P\in\mathbb{LR}$, \ $P=Q/Z^{\vec{p}}$, where \ $Q$ is a square-free polynomial, \ $\vec{p}= <p_1, \cdots, p_s>$ is a tuple of integers and \ $Z^{\vec{p}}=z_1^{p_1}\dots z_s^{p_s}$. \ $\{u_1, \cdots , u_r\}$, \ $\{v_1, \cdots , v_s\}$ are two real algebraic sets and linearly independent over $\mathbb{Q}$, then \ $\operatorname{LRhom}[u_1, \cdots, u_r, v_1, \cdots, v_s](P)$ has no multiple roots other than $0$.
  \begin{proof}
  Assume that \ $x_0\neq 0$ is a multiple root of \ $P_1(x)=\operatorname{LRhom}[u_1, \cdots, u_r, v_1, \cdots, v_s](P)$(short for \ $\operatorname{LRhom}(P)$, the same below), i.e. \ $P_1(x_0)=0$ and $P_1'(x_0)=0$.

   $\operatorname{LRhom}(P) = \operatorname{LRhom}(Q/Z^{\vec{p}}) = \operatorname{LRhom}(Q)/\operatorname{LRhom}(Z^{\vec{p}})$, let \ $Q_1(x)= \operatorname{LRhom}(Q)$ and \ $Z_P(x)=  \operatorname{LRhom}(Z^{\vec{p}})$, as \ $Z^{\vec{p}}=z_1^{p_1} \cdots z_s ^{p_s}$, \ $Z_P(x)= e^{I v_1 p_1 x} \cdots e^{I v_s p_s x}$, so \ $Z_P(x)\neq 0$ and \ $Z_P'(x)\neq 0$ for \ $\forall x\in \mathbb{R}$.

It is clear that  \ $P_1(x_0)=0$ implies that \ $Q_1(x_0)= 0$. As \ $P_1'(x)=(Q_1'(x) Z_P(x)- Q_1(x) Z_P'(x))/  \\ Z_P(x)^2$, so $P_1'(x_0)=0$ and \ $Q_1(x_0)= 0$ implies that \ $Q_1'(x_0)=0$. We get that \ $x_0$ is a multiple root of \ $Q_1(x)$, which contracts Corollary 2.1.
  \end{proof}
\end{theorem}

Theorem 2.3 is also a direct inference of Proposition 5 in ref[8].

Let $\operatorname{numerator}(G(x))$ denote the numerator of \ $G(x)$.

\begin{corollary}
For $F(x)=f(x, e^{u_{1}x},\cdots, e^{u_{r}x}, \sin(v_{1}x),\cdots, \sin(v_{s}x), \cos(r_{1}x),\cdots, \cos(r_{t}x))$, where \ $f\in\mathbb{C}[x, y_1, \cdots , y_r, z_{11},\cdots, z_{1s}, z_{21}, \cdots, z_{2t}]$, \ $u_{1}, \cdots, u_{r}, v_1, \cdots, v_{s}\in \mathbb{R}_{alg}$, let \ $G(x)=f(x,e^{u_{1}x},\cdots, \\ e^{u_{r}x}, \frac{e^{I v_{1}x}- e^{-I v_{1}x}}{2 I},\cdots,\frac{e^{I v_{1}x}- e^{-I v_{s}x}}{2 I},\frac{e^{I r_1 x}+e^{-I r_1 x}}{2},\cdots,\frac{e^{I r_t x}+e^{-I r_t x}}{2})$,  $A=(a_1,\cdots,a_m)$ an integral basis of $\{u_1, \cdots , \\ u_s\}$, $ B=(b_1,\cdots,b_n)$ an integral basis of \ $\{v_1, \cdots , v_s\, r_1, \cdots , r_s\}$, if the polynomial $hom[a_1,\cdots, a_m, \\ I b_1, \cdots,I b_n]^{-1} (\operatorname{numerator}(G(x))$ is square-free, then \ $F(x)$ has no multiple roots other than $0$.
\end{corollary}

We extend the operation of complex conjugation to \ $\mathbb{LR}$ as follows.

Given \ $P = \sum_{j=1}^{n}a_j(x) y_{1}^{v_{j1}}\cdots y_{r}^{v_{jr}} z_{1}^{w_{j1}}\cdots z_s^{w_{js}} \in  \mathbb{LR}$, where \ $a_1(x), \cdots, a_n (x) \in \mathbb{C}[x]$, define its formal conjugate to be \ $con(P) = \sum_{j=1}^{n}\overline{a_j(x)}y_{1}^{v_{j1}}\cdots y_{r}^{v_{jr}} z_{1}^{-w_{j1}}\cdots z_s^{-w_{js}}$, where \ $\overline{a_j(x)}$ is the conjugate of the complex function \ $a_j(x)$, and the same below. Obviously, \ $con(Z^{\vec{p}})=Z^{-\vec{p}}$.

\begin{lemma}\label{lemma2.5}
  For \ $P=\sum_{j=1}^{n}a_j(x) y_{1}^{v_{j1}}\cdots y_{r}^{v_{jr}} z_{1}^{w_{j1}}\cdots z_s^{w_{js}} \in  \mathbb{LR}$ and two arbitrary real algebraic numbers sets \ $\{u_1,\cdots,u_r\}$ and \ $\{v_1,\cdots,v_s\}$, \ $\operatorname{LRhom}(con( P))=\overline{\operatorname{LRhom}(P)}$.
  \begin{proof}
    \ $\operatorname{LRhom}(con(P))=\operatorname{LRhom}(\sum_{j=1}^{n}\overline{a_j(x)} y_{1}^{v_{j1}}\cdots y_{r}^{v_{jr}} z_{1}^{-w_{j1}}\cdots z_s^{-w_{js}})$

    \ $=\sum_{j=1}^{n}\overline{a_j(x)} y_{1}^{v_{j1}}\cdots y_{r}^{v_{jr}}(\cos(-w_{j_{1}}v_{1}x)+I\sin(-w_{j1}v_{1}x))\cdots (\cos(-w_{js}v_{s}x)+I\sin(-w_{js}v_{s}x))$

    \ $=\sum_{j=1}^{n}\overline{a_j(x)} y_{1}^{v_{j1}}\cdots y_{r}^{v_{jr}}(\cos(w_{j_{1}}v_{1}x)-I\sin(w_{j1}v_{1}x))\cdots (\cos(w_{js}v_{s}x)-I\sin(w_{js}v_{s}x))$

    \ $=\sum_{j=1}^{n}\overline{a_j(x)} y_{1}^{v_{j1}}\cdots y_{r}^{v_{jr}}\overline{(\cos(w_{j_{1}}v_{1}x)+I\sin(w_{j1}v_{1}x))}\cdots \overline{(\cos(w_{js}v_{s}x)-I\sin(w_{js}v_{s}x))}$

     \ $=\sum_{j=1}^{n}\overline{a_j(x) y_{1}^{v_{j1}}\cdots y_{r}^{v_{jr}} (\cos(w_{j_{1}}v_{1}x)+I\sin(w_{j1}v_{1}x))\cdots  (\cos(w_{js}v_{s}x)-I\sin(w_{js}v_{s}x))}$

    \ $=\overline{\operatorname{LRhom}(P)}$, \\where \ $y_{1}^{v_{j1}}=e^{v_{ji} u_i x}$ is real-valued for \ $i=1,\cdots, r$.
    \end{proof}
\end{lemma}

For \ $P=con(P)$, \ $\overline{\operatorname{LRhom}(P)}=\operatorname{LRhom}(con(P))=\operatorname{LRhom}(P)$ holds due to Lemma 2.5, so we get Lemma 2.6.

\begin{lemma}\label{lemma2.6}
For $P\in \mathbb{LR}$, if $P=con(P)$, then $\operatorname{LRhom}( P)$ is real-valued.
\end{lemma}

\begin{lemma}
For \ $P,P_1 , P_2\in \mathbb{LR}$, \ $con(P_1+P_2)=con(P_1)+con(P_2)$, \ $con(P_{1}P_2)=con(P_1)con(P_2)$, \ $con(con(P))=P$.
\end{lemma}

\begin{theorem}\label{theorem2.4}
  If \ $P = \sum_{j=1}^{n}a_j(x)x^{u_j}  y_{1}^{v_{j1}}\cdots y_{r}^{v_{jr}} z_{1}^{w_{j1}}\cdots z_s^{w_{js}} \in \mathbb{LR}$ such that \ $P=con(P)$, \ $P$ can be factorized as \ $C Z^{\vec{p}} P_1^{r_1} \cdots P_n^{r_n}$, where \ $C\in \mathbb{C}$ and \ $P_1,\cdots,P_n\in \mathbb{C}[x,y_1,\cdots,y_r,z_1,\cdots,z_s]$ are square-free and pairwisely co-prime, \ $\vec{p}=<p_1,\dots,p_s>$ is a tuple of integers and \ $Z^{\vec{p}}=z_1^{p_1}\dots z_s^{p_s}$, then for each \ $i$,

  1) there exist \ $C_i\in \mathbb{C}$,\ $\vec{p_i}$, a tuple of rational numbers, such that \ $P_i=C_i Z^{\vec{p_i}} con(P_i)$;

  2) For two arbitrary real algebraic numbers sets \ $\{u_1,\cdots,u_r\}$ and \ $\{v_1,\cdots,v_s\}$, \ $f_i(x)=\operatorname{LRhom}[u_1, \cdots, u_r; v_1,\cdots,v_s](P_i Z^{-\vec{p_i}/2} C_i^{-1/2})$ is real-valued or pure imaginary, furthermore, \ $f_i$ has no multiple root other than \ $0$ for \ $i=1,\cdots, n$,\ $f_i$ and \ $f_j$ have no common root other than \ $0$ for \ $i\neq j$;

  3) \ $\operatorname{LRhom}[u_1, \cdots, u_r; v_1,\cdots,v_s](P) = C_0 f_1^{r_1} \cdots f_n^{r_n}$, where \ $C_0 = C (C_1)^{r_1/2} \cdots (C_n)^{r_n/2}$.

  \begin{proof}
  1)For \ $n=1$, \ $P=con(P)$ implies \ $C Z^{\vec{p}} (P_1)^{r_1}= \overline{C} Z^{-\vec{p}} con(P_1)^{r_1}$, thus \ $P_1 = (\overline{C}/C)^{1/r_1} Z^{-2 \vec{p}/r_1} \\ con(P_1)$, the conclusion holds.

  Suppose the conclusion holds for \ $n=m-1$, that is to say, there exist $C_i$, $\vec{p_i}$ such that \ $P_i=C_i Z^{\vec{p_i}} con(P_i)$ for \ $1<i<m$.  Let \ $n=m$,

  $P=C Z^{\vec{p}} (P_1)^{r_1} \cdots (P_{m-1})^{r_{m-1}} P_m^{r_m}$

  $=C Z^{\vec{p}} (C_1)^{r_1} Z^{r_1 \vec{p_1}} con(P_1)^{r_1} \cdots (C_{m-1})^{r_{m-1}} Z^{r_{m-1} \vec{p_{m-1}}} con(P_{m-1})^{r_{m-1}} (P_m)^{r_m}$

  $=C_0 Z^{\vec{p_0}} con(P_1)^{r_1} \cdots con(P_{m-1})^{r_{m-1}} P_m^{r_m}$, \\
  where \ $C_0=C (C_1)^{r_1} \cdots (C_{m-1})^{r_{m-1}}$, \ $\vec{p_0}=\vec{p} + r_1 \vec{p_1} + \cdots+r_{m-1}\vec{p_{m-1}}$.

  On the other hand, \ $con(P)=\overline{C} Z^{-\vec{p}} con(P_1)^{r_1} \cdots con(P_{m-1})^{r_{m-1}} con(P_m)^{r_m}$.

  So, by the assumption \ $P = con(P)$, we get that if \ $con(P_1)\neq 0, \cdots,con(P_{m-1})\neq 0$, \ $P_m=C_m Z^{\vec{p_m}} con(P_m)$, where \ $C_m=(\overline{C}/C_0)^{1/r_m}$,\ $\vec{p_m}=(-\vec{p}-\vec{p_0})/r_m$. As \ $con(P_1),\cdots,con(P_{m-1})$ have finite zeros at most, \ $P_m$ and \ $con(P_m)$ are both continuous, \ so $P_m=C_m Z^{\vec{p_m}} con(P_m)$ holds.

  2)\ $P_i=C_i Z^{\vec{p_i}} con(P_i)$ implies \ $P_i^2=C_i Z^{\vec{p_i}} con(P_i) P_i$, so \ $P_i^2 C_i^{-1} Z^{-\vec{p_i}}=con(P_i) P_i$. Let \ $Q_i=P_i Z^{-\vec{p_i}/2} C_i^{-1/2}$, then \ $Q_i^2=con(P_i) P_i$, so that $Q_i^2=con(Q_i^2)$.

  Let \ $f_i(x) = \operatorname{LRhom}[u_1, \cdots, u_r; v_1,\cdots,v_s](Q_i)$, then \ $f_i(x)^2$ is real-valued due to Lemma 2.6, i.e. \ $f_i(x)$ is real-valued or pure imaginary.

  It is clearly that \ $f_i$ has no multiple root other than \ $0$ by Theorem 2.3,\ $f_i$ and \ $f_j$ have no common root other than $0$ for \ $i\neq j$ by Corollary 2.3.

  3)\ As \ $P=con(P)$, \ $degree(P, z_j)=degree(P,z_j^{-1})$ for \ $j=1,\cdots, s$,  denoted by \ $q_j$. It is trivial that \ $\vec{p}[j]=-q_j$ and  \ $degree(Q = P_1^{r_1} \cdots P_n^{r_n}, z_j)=2 q_j$. Let \ $p_{ij} = degree(P_i, z_j)=degree(con(P_i), z_j^{-1})$, then \ $\vec{p_i}=<p_{i1},\dots,p_{is}>$ for \ $i=1,\dots,n$ and \ $r_1 p_{1j}+\cdots+ r_n p_{nj} = degree(Q,z_j)= 2 q_j$ for \ $j=1,\dots,s$, so \ $\vec{p}[j]=-q_j=-(r_1 p_{1j}+\cdots+ r_n p_{nj})/2$ and \ $\vec{p}=-(r_1 \vec{p_{1}}+\cdots+ r_n \vec{p_{n}})/2$.

  So, \ $ C_0\times f_1^{r_1} \cdots f_n^{r_n} = C (C_1)^{r_1/2} \cdots (C_n)^{r_n/2} (\operatorname{LRhom}[u_1, \cdots, u_r; v_1,\cdots,v_s](P_1 Z^{-\vec{p_1}/2} C_1^{-1/2}))^{r_1} \cdots \\ (\operatorname{LRhom}[u_1, \cdots, u_r; v_1,\cdots,v_s](P_n Z^{-\vec{p_n}/2} C_n^{-1/2}))^{r_n}$

  \ \ $=  C\times (\operatorname{LRhom}[u_1, \cdots, u_r; v_1,\cdots,v_s](P_1 Z^{-\vec{p_1}/2} ))^{r_1} \cdots (\operatorname{LRhom}[u_1, \cdots, u_r; v_1,\cdots,v_s](P_n Z^{-\vec{p_n}/2}))^{r_n}$

  \ \ $= C\times \operatorname{LRhom}[u_1, \cdots, u_r; v_1,\cdots,v_s](Z^{-r_1\vec{p_1}/2 - \cdots - r_n \vec{p_n}/2} P_1^{r_1} \cdots P_n ^{r_n})$

  \ \ $=\operatorname{LRhom}[u_1, \cdots, u_r; v_1,\cdots,v_s](C Z^{\vec{p}} P_1^{r_1} \cdots P_n ^{r_n}) = \operatorname{LRhom}[u_1, \cdots, u_r; v_1,\cdots,v_s](P)$.
  \end{proof}
\end{theorem}

\begin{corollary}\label{corollary2.4}
  For each trigonometric-exponential polynomial \ $F(x)=f(x,e^{{u_1} x}, \cdots, e^{{u_r} x}, \sin(v_1 x), \\ \cdots, \sin(v_s x), \cos(v_1 x), \cdots, \cos(v_s x))$, where \ $f\in \mathbb{R}_{alg}[x, y_1, \cdots, y_r, z_1, \cdots, z_s, w_1, \cdots, w_s]$, there exists \ $C\in \mathbb{R}$ and real-valued  trigonometric-exponential polynomials \ $\{f_1,\dots, f_n\}$ such that \ $f_i$ has no multiple root other than $0$, $f_i$ and \ $f_j$ have no common root other than $0$ for \ $i\neq j$, \ $F(x)= C f_1(x)^{r_1} \cdots f_n(x)^{r_n}$.
\begin{proof}
Making substitution for F(x), \ $e^{{u_i} x}= y_i$ for \ $i=1, \cdots, r$, \ $\sin(v_i x)= \frac{z_i - 1/z_i}{2 I}$, \ $\cos(v_i x)= \frac{z_i + 1/z_i}{2}$ for \ $i=1, \cdots, s$, yields \ $P \in \mathbb{LR}$ such that \ $\operatorname{LRhom}[u_1, \cdots, u_r; v_1,\cdots,v_s](P)=F(x)$.

By Theorem 2.4, there exists \ $C\in \mathbb{C}$ and  trigonometric-exponential polynomials \ $\{f_i\}$ such that \ $f_i$ has no multiple root other than \ $0$ ,\ $f_i$ and \ $f_j$ have no common root other than \ $0$ for \ $i\neq j$, \ $F(x)= C f_1(x)^{r_1} \cdots f_n(x)^{r_n}$, each \ $f_i$ is real-valued or pure imaginary.

If all \ $f_i$s are real-valued, then Corollary 2.4 holds. For each \ $i$, if \ $f_i$ is pure imaginary, let \ $f_i'=\frac{f_i}{I}$ and \ $C'=C I^{r_i}$, then  $f_i'$ is real-valued and \ $F(x)= C' f_1(x)^{r_1} \cdots {f_i'(x)}^{r_i} \cdots f_n(x)^{r_n}$.

Now, \ $F(x)$ and all \ $f_i(x) (or f_i'(x))$s are real-valued, so the constant \ $C$ (or $C'$) must be a real number. That is to say the corollary holds.

\end{proof}
\end{corollary}

\begin{algorithm}[H]
\caption{algorithm caption}
\LinesNumbered 
\KwIn{A TEP $F(x)=f(x,e^{u_{1}x},\cdots,e^{u_{r}x},\sin(v_{1}x),\cdots,\sin(v_{s}x),\cos(r_{1}x),\cdots,\cos(r_{t}x))$, where \ $f\in \mathbb{R}_{alg}[x,y_1,\cdots,y_{r},z_{1},\cdots,z_{s},w_{1},\cdots,w_{t}]$}
\KwOut{$F(x)=C F_1(x)^{n_1}\cdots F_m(x)^{n_m}$ such that $C\in \mathbb{R}_{alg}$ and \ $F_i(x)$ is a real-valued TEP which has no multiple roots other than $0$  for $i=1, \cdots , m$, $F_i(x)$ and $F_j(x)$ have no common roots other than $0$ for $i\neq j$}

$A=\{a_1,\cdots,a_m\} \leftarrow $ an integral basis of $\{u_1, \cdots , u_r\}$

$B=\{b_1,\cdots,b_n\} \leftarrow $ an integral basis of $\{v_1, \cdots , v_s , r_1, \cdots , r_t\}$ \;

$G\leftarrow f(x,e^{u_{1}x},\cdots,e^{u_{r}x},\frac{e^{I v_{1}x}- e^{-I v_{1}x}}{2 I},\cdots,\frac{e^{I v_{1}x}- e^{-I v_{s}x}}{2 I},\frac{e^{I r_1 x}+e^{-I r_1 x}}{2},\cdots,\frac{e^{I r_t x}+e^{-I r_t x}}{2})$

  $G_1\leftarrow numerator(G)$

  $G_2\leftarrow denominator(G)$

  $Q_1\leftarrow hom[a_1,\cdots, a_m,I b_1,\cdots, I b_n]^{-1}(G_1)$, where $Q_1(x_1, y_1,\cdots, y_m, z_1,\cdots, z_n)$ is an \ $(m+n+1)$-ary polynomial and \ $y_i=e^{a_i x}$, $z_j=e^{I b_j x}$ for $i=1, \cdots, m$, $j=1, \cdots, n$

  $Q_2\leftarrow hom[a_1,\cdots, a_m, I b_1,\cdots,I b_n]^{-1}(G_2)$, where $Q_2=Z^{\vec{p}}=z_1^{p_1} \cdots z_n^{p_n}$\;

  $P\_lst\gets factor(Q_1)=C P_{1}^{r_1}\cdots P_n^{r_n}$, where \ $P_1, \cdots, P_n$ are square-free and pairwise co-prime

  $P\_lst\gets Z^{-\vec{p}} P\_lst$; \ i.e. $ P\_lst= Z^{-\vec{p}} factor(Q_1)=C Z^{-\vec{p}} P_{1}^{r_1}\cdots P_n^{r_n}$

  $F\gets1$

  $C_0\gets C$

\For {$i$ form $1$ to $n$}{

  $g\gets P_i$

  $h\gets g/con(g)$, where $h$ is of form \ $C_{i} z_{1}^{-p_1} \cdots z_{n}^{-p_n}$, \ $LRhom(g^2/h)=LRhom(g \times con(g))$ is real-valued

  $f_i\gets \operatorname{LRhom}[a_1,\cdots, a_m, b_1,\cdots, b_n](g/(C_i^{\frac{1}{2}}z_1^{-\frac{p_1}{2}} \cdots z_{n}^{-\frac{p_{n}}{2}}))$, where $f_i$ is real-valued or pure imaginary by Theorem 2.4.

\If{$f_i$ is pure imaginary}{

  $f_i\gets \frac{f_i}{I}$

  $C_0\gets C_{0} I^{r_i}$}

  $F\gets F\times f_{i}^{r_i}$

  $C_0\gets C_{0}\times(C_{i}^{1/2})^{r_i}$}

  \Return $C_{0}\times F$

\end{algorithm}

The following example shows the process of Algorithm 1.

\begin{example}
  Factorize the trigonometric polynomial  $F(x)=1-\cos^3(x)-\sin^3(x)$.

    Let $z=e^{Ix}$, so that \ $\cos(x)=\frac{z + z^{-1}}{2}$ and \ $\sin(x)=\frac{z - z^{-1}}{2 I}$ , then we can get a Laurent polynomial of \ $F(x)$ on complex field, \ $P = 1-(\frac{z + z^{-1}}{2})^3-(\frac{z - z^{-1}}{2 I})^3$.

   Factorization yields that \ $P= \frac{(-\frac{1}{8}-\frac{I}{8})(z^2+2z+2Iz+I)(z-1)^2(-z+I)^2}{z^3}$.

Let $P_1=z^2+2z+2Iz+I$, then \ $con(P_1)=\frac{1}{z^2}+\frac{2}{z}-\frac{2I}{z}-I$, and $P_1=c_{1} z^{2} con(P_1)$, where $c_1=1$.

\begin{equation}
\begin{aligned}
f_1 &=\operatorname{LRhom}[; 1](P_1/(I^{\frac{1}{2}} z))\\
&=\operatorname{LRhom}[; 1](I^{-\frac{1}{2}}z+2I^{-\frac{1}{2}}+2I^{\frac{1}{2}}+I^{\frac{1}{2}}z^{-1})\\
&=e^{-I \frac{\pi}{4}+I x}+2 e^{-I\frac{\pi}{4}} + 2e^{I\frac{\pi}{4}}+e^{I\frac{\pi}{4}-Ix}\\
&=\cos(x-\frac{\pi}{4})+I\sin(x-\frac{\pi}{4})+2(\cos(-\frac{\pi}{4})+I\sin(-\frac{\pi}{4}))+ 2(\cos(\frac{\pi}{4})+I\sin(\frac{\pi}{4}))+ \\
&\cos(x-\frac{\pi}{4})-I\sin(x-\frac{\pi}{4}) \\
&=\cos(x-\frac{\pi}{4})+4\cos(\frac{\pi}{4})\\
&=\sqrt{2}(\cos(x)+\sin(x)+2)
\nonumber
\end{aligned}
\end{equation}

Let \ $P_2=z-1$ , \ $con(P_2)=\frac{1}{z}-1$ , \ $P_2=c_{2} z con(P_2)$, where \ $c_2=-1=e^{I \pi}$.

\begin{equation}
\begin{aligned}
f_2 &=\operatorname{LRhom}[; 1](P_2/(c_2 z)^{\frac{1}{2}})=\operatorname{LRhom}[; 1](z^{\frac{1}{2}}-1/(c_2 z)^{\frac{1}{2}})\\
&=e^{I (\frac{x}{2}-\frac{\pi}{2})}-e^{-I(\frac{x}{2}+\frac{\pi}{2})} \\
&=\cos(\frac{x}{2}-\frac{\pi}{2})+I\sin(\frac{x}{2}-\frac{\pi}{2})-(\cos(\frac{x}{2}+\frac{\pi}{2})-I\sin(\frac{x}{2}+\frac{\pi}{2}))\\
&= \sin(\frac{x}{2})-I \cos(\frac{x}{2}) -(-\sin(\frac{x}{2}) -I \cos(\frac{x}{2})) = 2\sin(\frac{x}{2}).
\nonumber
\end{aligned}
\end{equation}

Let \ $P_3=-z+I$, \ $con(P_3)=-\frac{1}{z}-I$, $P_3=c_3 z con(P_3)$, where \ $c_3=-I=e^{- I \frac{\pi}{2}}$,
\begin{equation}
\begin{aligned}
f_3 &=\operatorname{LRhom}[; 1](P_3/(-I z)^{\frac{1}{2}}=\operatorname{LRhom}[; 1](-z+I)/(-I z)^{\frac{1}{2}}\\
&=(-e^{I x}+e^{I \frac{\pi}{2}})/e^{-I \frac{\pi}{4}+I \frac{x}{2}}
=-e^{I \frac{x}{2}+I \frac{\pi}{4}}+e^{I \frac{3\pi}{4}-I \frac{x}{2}}\\
&=-\cos(\frac{x}{2}+\frac{\pi}{4})-I\sin(\frac{x}{2}+\frac{\pi}{4})+\cos(\frac{3\pi}{4}-\frac{x}{2})+I\sin(\frac{3\pi}{4}-\frac{x}{2})\\
&=-\cos(\frac{x}{2}+\frac{\pi}{4})-I\sin(\frac{x}{2}+\frac{\pi}{4})-\cos(\frac{\pi}{4}+\frac{x}{2})+I\sin(\frac{\pi}{4}+\frac{x}{2})\\
&=-2\cos(\frac{x}{2}+\frac{\pi}{4})=-\sqrt{2}(\cos(\frac{x}{2})-\sin(\frac{x}{2}))=\sqrt{2}(\sin(\frac{x}{2})-\cos(\frac{x}{2})),
\nonumber
\end{aligned}
\end{equation}

  \begin{equation}
  \begin{aligned}
  C_0 &=C c_1^{\frac{1}{2}} c_2 c_3=(-\frac{1}{8}-\frac{I}{8}) (I^{\frac{1}{2}}) (-1) (-I)=-(\frac{1}{8}+\frac{I}{8}) I^{\frac{3}{2}}\\
  &=\frac{(1+I) I^{-\frac{1}{2}}}{8}=\frac{I^{-\frac{1}{2}}+I^{\frac{1}{2}}}{8} =\frac{e^{-I \frac{\pi}{4}}+e^{I \frac{\pi}{4}}}{8}\\
  &=\frac{\cos(\frac{\pi}{4})-I\sin(\frac{\pi}{4})+\cos(\frac{\pi}{4})+I\sin(\frac{\pi}{4})}{8} \\
   &=-\frac{\sqrt{2}}{8}.
   \nonumber
  \end{aligned}
  \end{equation}

Therefore, $F(x)=C_0\times f_1\times f_2^{2}\times f_3^{2}$, where \ $C_0= -\frac{\sqrt{2}}{8}$, $f_1=\sqrt{2}(\cos(x)+\sin(x)+2)$, $f_2=2 \sin(\frac{x}{2})$, $f_3=\sqrt{2}(\sin(\frac{x}{2})-\cos(\frac{x}{2}))$, and \ $f_1$,\ $f_2$,\ $f_3$ are pairwise co-prime and have no multiple roots, i.e. \ $f_1 \times f_2 \times f_3$ have no multiple roots.

\end{example}

\begin{example}
Decide whether trigonometric-exponential polynomial $f(x)=-\sqrt{3}-24\sqrt{3} e^{-x}-4\sin(\frac{\sqrt{3}}{2}x) e^{-\frac{3}{2}x}-12\sqrt{3} e^{-\frac{5}{2}x}+108\sqrt{3} e^{-2x}-8e^{-3x} \sin(\frac{\sqrt{3}}{2}x)\cos(\frac{\sqrt{3}}{2}x)+36e^{-\frac{5}{2} x}\sin(\frac{\sqrt{3}}{2}x)$ has real multiple roots.

As \ $\{-1,-\frac{3}{2},-\frac{5}{2},-2,-3,-\frac{5}{2}\}$ has an integral basis \ $\{-\frac{1}{2}\}$, let \ $y_1=e^{-\frac{1}{2} x}$ and \ $z_1=e^{I \frac{\sqrt{3}}{2}x}$, then $f(x)$ can be transformed to a Laurent polynomial \ $P(x,y_1,z_1) = -\sqrt{3}-24\sqrt{3} y_1^2- 4\times \frac{z_1-\frac{1}{z_1}}{2 I}\times y_1^3-12\sqrt{3} y_1^5+108\sqrt{3} y_1^4-8y_1^6\times \frac{z_1-\frac{1}{z_1}}{2 I}\times \frac{z_1+\frac{1}{z_1}}{2} + 36y_1^5\times\frac{z_1-\frac{1}{z_1}}{2 I}$.

\ $factor(P)=-\frac{1}{7} \times \frac{1}{z_1^2} \times (\sqrt{3}-2 I)\times (-y_1^6-(6 I) \sqrt{3} y_1^5 z_1-(2 I) \sqrt{3} z_1^2-(2 I) \sqrt{3} y_1^3 z_1^3+(2 I) \sqrt{3} y_1^3 z_1-(216 I) \sqrt{3} y_1^4 z_1^2+(48 I) \sqrt{3} y_1^2 z_1^2+(4 I) \sqrt{3} y_1^6-3 z_1^2+72 y_1^2 z_1^2+7 y_1^6 z_1^4+6 y_1^6 z_1^2-18 y_1^5 z_1^3+54 y_1^5 z_1-324 y_1^4 z_1^2+(4 I) \sqrt{3} y_1^6 z_1^2+(30 I) \sqrt{3} y_1^5 z_1^3+4 y_1^3 z_1^3-4 y_1^3 z_1)$.

So \ $factor(P)$ can be wrote as \ $-\frac{1}{7} \times \frac{1}{z_1^2} \times Q$, where \ $Q$ is square-free, then we get that \ $f(x)$ has no real multiple roots by Theorem 2.3.
\end{example}

\section{Succusive Taylor Substitution}

In this section, we are to discuss the sign-deciding of transcendental polynomials by Successive Taylor-substitution.

\begin{definition}
For a transcendental function \ $F(x)$, on a certain interval \ $I$, if there are two algebraic functions sequences \ $\{\operatorname{T_{min}}(n,F)\}$ and \ $\{\operatorname{T_{max}}(n,F)\}$, and an \ $n_0 \in \mathbb{N}$ such that for \ $n \geq n_0$,

1)\ $\operatorname{T_{min}}(n+1,F)> \operatorname{T_{min}}(n,F)$, and for \ $n\rightarrow \infty$,\ $\operatorname{T_{min}}(n,F)\rightarrow F(x)$,

2)\ $\operatorname{T_{max}}(n+1,F)< \operatorname{T_{max}}(n,F)$, and for \ $n\rightarrow \infty$,\ $\operatorname{T_{max}}(n,F)\rightarrow F(x)$.

We call \ $\{\operatorname{T_{min}}(n,F)\}$ and \ $\{\operatorname{T_{max}}(n,F)\}$ the lower limit polynomials sequence and upper limit polynomials sequence of \ $F(x)$ on interval \ $I$ respectively, \ $\operatorname{T_{min}}(n,F)$ is the lower limit polynomial of \ $F(x)$, and \ $\operatorname{T_{max}}(n,F)$ is the upper limit of \ $F(x)$. \ $n_0$ is called the threshold.
\end{definition}

Obviously, the lower limit polynomials and the upper limit polynomials of \ $F(x)$ satisfy:

$\operatorname{T_{min}}(n_0,F)(x)<\operatorname{T_{min}}(n_{0}+1,F)(x)<\operatorname{T_{min}}(n_{0}+2,F)(x)<\cdots<F(x)<\cdots<\operatorname{T_{max}}(n_{0}+2,F)(x)< \operatorname{T_{max}}(n_{0}+1,F)(x)< \operatorname{T_{max}}(n_{0},F)(x)$.

Therefore, there are nested polynomials to approximate \ $F(x)$:

\ $(\operatorname{T_{min}}(n_0,F)(x),\operatorname{T_{max}}(n_0,F)(x))\supset (\operatorname{T_{min}}(n_{0}+1,F)(x),\operatorname{T_{max}}(n_0+1,F)(x))\supset(\operatorname{T_{min}}(n_{0}\\ + 2,F)(x),\operatorname{T_{max}}(n_{0}+2, F)(x))\supset \cdots \supset \{F(x)\}$.

Next, we are to discuss how to obtain the lower limit polynomials and upper limit polynomials for a specific class of transcendental polynomial with the form \ $F(x)=f(x,trans(x))$, where \ $f(x,y)$ is a binary polynomial.

For convenience, we take the sum of the first \ $n$ terms in Taylor expansion of function \ $f(x)$ at \ $0$ as \ $taylor(f,n)$. Obviously, if \ $taylor(f,n)$ converges to \ $f(x)$, \ $f(x) = Taylor(f, n)+o(x^p)$, where \ $p\geq n$.

\begin{definition}
  If the transcendental function \ $f(x)$ meets the following conditions on a certain intervals \ $[0,T]$,

  1)\ For \ $x\in [0,T]$,\ $f(x)\geq 0$ and for \ $x\in (0, T]$, \ $f(x)> 0$;

  2)\ Taylor expansion of \ $f(x)$ at \ $0$ is a staggered series and converges to \ $f(x)$, i.e. \ $taylor(f,n)=\sum_{i=1}^{n}(-1)^{i-1}f_i(x)$, 
where \ $f_i(x)=a_{i}x^{m_i}$, \ $0 < a_i \leq 1$, \ $m_{i-1} < m_i$;

  3)\ there exists a constant \ $n_0(f,T)$, for \ $n\geq n_0(f,T)$, \ $taylor(f,n)>0$ and \ $f_n(x) > f_{n+1}(x)>0$.

   We say that \ $f(x)$ can be regularly expanded on the interval \ $[0, T]$, and the constant \ $n_0(f,T)$ is called its threshold.
\end{definition}

Most of the common basic elementary transcendental functions can be regularly expanded on the corresponding intervals.

\ $\arctan(x)= x - \frac{x^3}{3} + \frac{x^5}{5} - \frac{x^7}{7} +\cdots+ \frac{(-1)^{n-1}x^{2n-1}}{2n-1} +\cdots $(\ $|x|\leq 1$), can be regularly expanded on \ $(0,1]$, and the threshold \ $n_0=1$.

\ $e^{-x}= 1 - x + \frac{x^2}{2!} - \frac{x^3}{3!} +\cdots + \frac{(-x)^{n-1}}{(n-1)!}+\cdots$, can be regularly expanded on \ $(0,T]$, where \ $T>0$ and the threshold \ $n_0= \min\{n\in \mathbb{N},for \ 0<x\leq T, taylor(e^{-x}, n)>0\ and\  taylor(e^{-x}, n+1)>0\}$.

$\ln(1+x)= x - \frac{x^2}{2} + \frac{x^3}{3} -\cdots+ \frac{(-1)^{k-1}x^k}{k}+\cdots(0<x<1)$ , can be regularly expanded on \ $(0,1)$, and the threshold \ $n_0=1$.

$\sin(x) = x- \frac{x^3}{3!} + \frac{x^5}{5!} -\cdots+ \frac{(-1)^{k-1}x^{2k-1}}{(2k-1)!} + \cdots $, can be regularly expanded on \ $(0,T]$, where \ $0<T<\pi$ and the threshold \ $n_0= \min\{n\in \mathbb{N}, for \ 0<x\leq T, taylor(\sin(x), n)>0 ,\ taylor(\sin(x), n+1)>0 \}$, e.g. for \ $T=\frac{\pi}{2}$, \ $n_0=1$; for \ $T=3$,\ $n_0=3$; for \ $T=\frac{314}{100}$, $n_0=5$.

\ $\cos(x) = 1 - \frac{x^2}{2!} + \frac{x^4}{4!} -\cdots+ \frac{(-1)^{k}x^{2k}}{(2k)!} +\cdots $, can be regularly expanded on \ $(0,T]$, where \ $0<T<\frac{\pi}{2}$ and the threshold \ $n_0=1$. When \ $x=\frac{\pi}{2}$ , as \ $taylor(\cos(x), 2n)<\cos(x)=0$, so \ $\cos(x)$ can not be regularly expanded on \ $(0,\frac{\pi}{2}]$.

There are also some elementary transcendental functions which can not be regularly expanded. e.g.

$\arcsin x = x + \frac{1}{2}\frac{x^3}{3} + \frac{1\times 3}{2\times 4}\frac{x^5}{5} + \cdots $ (\ $|x|<1$),

$e^x = 1+x+ \frac{x^2}{2!} + \frac{x^3}{3!} +\cdots + \frac{x^n}{n!} +\cdots$.

\begin{lemma}
  If \ $f(x)$ can be regularly expanded on given interval \ $I$ and the threshold is \ $n_0$, then for \ $n\geq n_0$, on \ $I$,

  1)\ $taylor(f, 2n-2))< taylor(f, 2n)< f(x)$,

  2)\ $taylor(f, 2n-1)> taylor(f, 2n+1)> f(x)$,

  3)\ when \ $n\rightarrow\infty, taylor(f,n)\rightarrow f(x)$.
\end{lemma}

Denote the sums of the positive and negative terms of the expansion of polynomial $f$ by $f^+$ and $f^{-}$ respectively. Obviously, \ $f= f^{+}+f^-$ and the following lemma clearly holds.

\begin{lemma}
If \ $T_1(y)>0$,\ $T_2(y)>0$ and \ $T_1(y)<x<T_2(y)$,\ then \ $f^+(T_1(y),y)+ f^{-}(T_2(y),y)<f(x,y)< f^{+}(T_2(y),y)+ f^{-}(T_1(y),y)$.
\end{lemma}

\begin{theorem}\label{theorem3.1}
  If \ $F(x)=f(x,trans(x))$ and \ $trans(x)$ can be regularly expanded on $I$, then

  \ $\operatorname{T_{max}}(n,F)= f^{+}(x,taylor(trans(x),2n-1))+f^{-}(x,taylor(trans(x),2n))$ is an upper limit polynomial of \ $F(x)$,

  \ $\operatorname{T_{min}}(n,F)= f^{+}(x,taylor(trans(x),2n))+f^{-}(x,taylor(trans(x),2n-1))$ is a lower limit polynomial of \ $F(x)$.
\end{theorem}

In the above scheme, the upper and lower polynomials sequences of transcendental function can be obtained by Taylor expansion, and a series of nested unary polynomials to approximate the objective function are established, so that the sign decision of the objective function can be transformed to a series of verifications of the unary polynomial inequalities,  and then the final work can be completed by means of algebraic inequality proving tools (such as xprove of BOTTEMA). We call this scheme Taylor-substitution.

A general transcendental polynomial may contain several transcendental factors, and the expression after Taylor substitution once may still contains transcendental factors, which need Taylor substitution again or even many times. The transcendental polynomial with one more transcendental factors with the form \ $f(x) = f(x, trans_1(x),\cdots, trans_n(x)))$ will be discussed below.

\begin{lemma}
For $f(x,y_1,\cdots,y_n)\in \mathbb{R}[x,y_1,\cdots,y_n]$, expressions $T_{1i}$ and $T_{2i}$ such that $T_{2i}>y_i>T_{1i}>0$ for \ $i=1,\cdots,n$, then \ $f^+ (x, T_{11}, T_{12},\cdots, T_{1n}) + f^-(x, T_{21}, T_{22},\cdots, T_{2n}) < f(x_1,\cdots, x_n) < f^{+}(T_{21}, T_{22},\cdots, T_{2n}) + f^{-}(x, T_{11}, T_{12},\cdots, T_{1n})$.
\end{lemma}

\begin{theorem}
For \ $f(x,y_1,\cdots,y_t)\in \mathbb{R}[x,y_1,\cdots, y_t]$, transcendental functions \ $trans_1(x),\cdots, \\ trans_t(x)$ can be regularly expanded on
 \ $(0,T)$,  the thresholds are \ $n_1,\cdots, n_t$ respectively. \ $n_0=\max\{n_1,\cdots, n_t\}$,\ $F(x)=f(x,trans_1(x),\cdots,trans_t(x))$, 
when \ $n\geq n_0$,

$\operatorname{T_{max}}(n,F) = f^+ (x,taylor(trans_1(x),2n-1), \cdots, taylor(trans_t(x),2n-1)) + f^- (x,taylor(trans_1(x), 2n), 
\cdots, taylor(trans_t(x),2n))$ is an upper limit polynomial of $F(x)$,

$\operatorname{T_{min}}(n,F) = f^+(x,taylor(trans_1(x),2n), \cdots, taylor(trans_t(x),2n)) + f^- (x,taylor(trans_1(x), 2n-1),
\cdots, taylor(trans_t(x),2n-1))$ is a lower limit polynomial of $F(x)$.

For \ $f(x,y_1,\cdots,y_t)=\sum_{i=1}^{s}f_i(x)\prod_{j=1}^{t}(y_j)^{d_{ij}}$,

 $\operatorname{T_{max}}=\sum_{i=1}^{s}(f_i^{+}(x)(\prod_{j=1}^{t}(taylor(trans_j(x), \\ 2n-1))^{d_{ij}}) +f_i^{-}(x)(\prod_{j=1}^{t}(taylor(trans_j(x),2n)^{d_{ij}}))$ is an upper limit polynomial of \ $F(x)$,

$\operatorname{T_{min}}(n,F)=\sum_{i=1}^{s}(f_i^{+}(x)(\prod_{j=1}^{t}(taylor(trans_j(x),2n))^{d_{ij}}) +f_i^{-}(x)(\prod_{j=1}^{t}(taylor(trans_j(x),2n-1))^{d_{ij}}))$ is a lower limit polynomial of \ $F(x)$.
\end{theorem}

We call the above scheme Successive Taylor-substitution. In the subsequent discussion, unless otherwise specified, all the upper and lower limit polynomials \ $\operatorname{T_{max}}(n,F)$ and \ $\operatorname{T_{min}}(n,F)$ of \ $F(x)$ refer to the definitions in Theorem 3.1 and Theorem 3.2. If transcendental function $F(x)$ has lower limit polynomials sequence and upper limit polynomials sequence, the decision of the sign of \ $F(x)$ can be fulfilled by the following Algorithm 2.

\begin{algorithm}[H]
\caption{Deciding\_transcendental\_polynomial}
\LinesNumbered 
\KwIn{$F (x) = f(x,trans_1(x),\cdots,trans_t(x))$ and an interval \ $I$}
\KwOut{The sign of \ $F(x)$ on $I$
 $n \leftarrow \operatorname{max} \{n_1,\cdots,n_t\}$, where \ $n_i$ is the threshold of \ $trans_i(x)$ for being regularly expanded; compute the upper limit polynomial \ $\operatorname{T_{max}}(n, f)$ and the lower limit polynomial \ $\operatorname{T_{min}}(n, f)$}

\If{$\operatorname{T_{min}}(n, f)\geq 0$ holds on $I$}{\Return{1}    \tcp{$F(x)> 0$ holds}
} 

\If{$\operatorname{T_{max}}(n, f)\leq 0$ holds on $I$}{\Return{-1}    \tcp{$F(x)< 0$ holds}
} 

\eIf{neither $\operatorname{T_{max}}(n,f)\geq 0$ nor $\operatorname{T_{min}}(n,f)\leq 0$ holds on \ $I$}{\Return{0}   \tcp{$F(x)$ has no constant sign on $I$}}{$n\leftarrow n+1$, goto 2)}

\end{algorithm}

[4,6] proved that for \ $trans(x)=\arctan(x) (0<x\leq 1)$ or $trans(x)=e^{-x}$, $f(x,y)\in \mathbb{Q}[x,y]$, Algorithm 2 is correct and will definitely terminate. It is also pointed that for more general transcendental polynomials, the conclusion about the root multiplicities of transcendental polynomials may not hold, and the algorithm may not terminate. Fortunately, Algorithm 1 in this paper can reduce the root multiplicities of TEP.

\begin{theorem}\label{theorem3.3}
$\{\operatorname{T_{min}}(n,F)\}$ and \ $\{\operatorname{T_{max}}(n,F)\}$ are the lower and upper polynomials sequences of transcendental polynomial \ $F(x)$ on the interval $I$, for any \ $x_0\in I$, \ $f(x_0)<0$ if and only if there exists \ $n_0\in \mathbb{N}$ such that \ $\operatorname{T_{max}}(n_0, F)(x_0)<0$,\ $f(x_0)>0$ if and only if there exists \ $n_0\in \mathbb{N}$ such that \ $\operatorname{T_{min}}(n_0, f)(x_0)>0$.
  \begin{proof}
We are to prove the former part firstly.

Denote the threshold for \ $\operatorname{T_{max}}(n,F)$ is the upper polynomial of $F(x)$ by \ $n(I)$. Obviously, for \ $n\geq n(I)$, \ $\operatorname{T_{max}}(n,F)>F$, so the sufficiency obviously holds.

Assume that for any \ $n\in \mathbb{N}$, \ $\operatorname{T_{max}}(n,F)(x_0)\geq 0$ holds, then \ $n\rightarrow \infty, \operatorname{T_{max}}(n,F)(x_0)\rightarrow F(x_0)$, we get that \ $F(x_0)\geq 0$, which contradicts the known conditions, so the necessity holds.

  In the same way, \ $F(x_0)>0$ if and only if there exists \ $n_0$ such that \ $\operatorname{T_{min}}(n_0,F)(x_0)>0$.
  \end{proof}

\end{theorem}

\begin{lemma}\label{lemma3.4}
If \ $trans_1(x),\cdots,trans_t(x)$ can be regularly expanded on \ $(0,T]$,\  $f(x,y_1,\cdots,y_t)\in \mathbb{R}[x,y_1,\cdots,y_t]$, \ $F(x)=f(x,trans_1(x),\cdots,trans_t(x))$, then there exists constants \ $M$ and \ $N$ independent of \ $n$ such that the sum of absolute values of all coefficients of the polynomial \ $\operatorname{T_{min}}(n, f)(x)$ after expansion is less than \ $M \times n^N$.

\begin{proof}
For a real polynomial \ $f$, denote the sum of the absolute value of the coefficients by \ $||f||$.

As for each \ $trans_i(x)$, the absolute value of each coefficient of \ $taylor(trans_i(x),n)$ is less than or equal to \ $1$, so \ $||taylor(trans_i(x),n)||\leq n$.

Let \ $f(x,y_1,\cdots,y_t)=\sum_{i=1}^{s}f_i(x)(\prod_{j=1}^{t}(y_j)^{d_{ij}})$, then \ $F(x)=\sum_{i=1}^{s}f_i(x)(\prod_{j=1}^{t}(trans_j(x))^{d_{ij}})$, thereby,

$\operatorname{T_{min}}(n,F)(x)=\sum_{i=1}^{s}(f_i^{+}(x)(\prod_{j=1}^{t}(taylor(trans_j(x),2n))^{d_{ij}}) +f_i^{-}(x)(\prod_{j=1}^{t}(taylor(trans_j(x),\\ 2n-1))^{d_{ij}}))$.

So, \ $||\operatorname{T_{min}}(n,F)||\\ \leq \sum_{i=1}^s(||f_i^{+}(x)||(\prod_{j=1}^{t}(||taylor(trans_j(x),2n)||)^{d_{ij}})+||f_{i}^{-}(x)||(\prod_{j=1}^t(||taylor(trans_j(x),2n-1)||)^{d_{ij}}))\\ \leq \sum_{i=1}^{s}(||f_{i}^{+}(x)||(\prod_{j=1}^{t}(2n)^{d_{ij}})+||f_i^{-}(x)||(\prod_{j=1}^{t}(2n-1)^{d_{ij}}))\\ <\sum_{i=1}^{s}(||f_{i}^{+}(x)||(\prod_{j=1}^{t}(2n)^{d_{ij}})+||f_{i}^{-}(x)||(\prod_{j=1}^t(2n)^{d_{ij}}))\\  \leq \sum_{i=1}^s(||f_i(x)||((2n)^{d_{i1}+\cdots+d_{it}})$.

Let \ $M_i=||f_i(x)||$,\ $N_i=d_{i1}+\cdots+d_{it}$,\ $M'=\max\{M_{i}\times 2^{N_i}|i=1\cdots s\}$, \ $N=\max\{N_i|i=1\cdots s\}$, then \ $||\operatorname{T_{min}}(n,F)||\leq M' \sum_{i=1}^{s}n^{N_i}\leq s\times M'\times n^N$.

  Let \ $M=s\times M'$, then  \ $||\operatorname{T_{min}}(n,F)||\leq M\times n^N$ and \ $M$ ,\ $N$ are independent of \ $n$.
  \end{proof}

\end{lemma}

Denote the lowest degree of the univariate polynomial \ $g$ by \ $td(g)$, and the coefficient of the lowest degree term by \ $tc(g)$.

\begin{lemma}\label{lemma3.5}
If transcendental functions \ $trans_1(x),\cdots,trans_t(x)$ can be regularly expanded on a given interval \ $(0,T]$, \ $f(x,y_1,\cdots,y_t)\in \mathbb{R}[x,y_1,\cdots,y_t]$, suppose that \ $F(x)=f(x,trans_1(x),\cdots,\\ trans_t(x))>0$ holds on \ $(0,T)$, then there exists \ $ n_0\in \mathbb{N}$ such that \ $td(\operatorname{T_{min}}(n_0,F)) <2n_0-1$.

\begin{proof}
For \ $n\in \mathbb{N}$, denote \ $d_n= td(\operatorname{T_{min}}(n,F))$, \ $c_n=tc(\operatorname{T_{min}}(n,F))$, that is to say, \ $\operatorname{T_{min}}(n,F)=c_{n}x^{d_n}+g(x)$,\ the degree of each item of \ $g(x)$ is bigger than \ $d_n$, by Lemma 3.4, the absolute value of  \ $c_n$ and each coefficient of \ $g(x)$ are less than \ $M\times n^N$, where \ $M$ and \ $N$ are constants independent of \ $n$.

Let \ $\lambda=\min\{T,1\}$, then for $0<x<\lambda$,\ $|\operatorname{T_{min}}(n,F)|<M\times n^{N}\times x^{d_n}\times (1+x+x^2+\cdots)<M\times n^N\times x^{d_n}/(1-x)$.

Assume that for any \ $n\in \mathbb{N}$,\ $d_n\geq 2n-1$, then \ $|\operatorname{T_{min}}(n,F)|\leq M\times n^N\times x^{2n-1}/(1-x)$. It is easy to prove that when \ $n\rightarrow\infty$, \ $\|\operatorname{T_{min}}(n, f )(x)\| \rightarrow 0$, which contradicts the fact that \ $\operatorname{T_{min}}(n, f)\rightarrow F$ and the assumed condition that
 \ $F(x)> 0$ on \ $(0,\lambda)\subset (0,T)$. So there exists \ $n_0\in \mathbb{N}$ such that \ $td(\operatorname{T_{min}}(n_0,f)) < 2n_0 - 1$.
 \end{proof}
\end{lemma}

Let \ $o(x^p)$ be the higher order infinitesimal of \ $x^p$ for \ $x\rightarrow 0$.

\begin{lemma}\label{lemma3.6}
  For \ $T\in (0,1)$, if \ $F(x)=f(x,trans_1(x),\cdots,trans_t(x))>0$ holds on \ $(0,T)$, then there exists \ $n_0\in \mathbb{N}$ and \ $\delta\in(0,T)$ such that for \ $x\in(0,\delta)$, \ $\operatorname{T_{min}}(n_0,F)(x)>0$.
  \begin{proof}
  \ $F(x)=\sum_{i=1}^{s}f_i(x)(\prod_{j=1}^{t}(trans_j(x))^{d_{ij}})\\
  \ \ \  =\sum_{i=1}^{s}(f_i^{+}(x)(\prod_{j=1}^{t}(taylor(trans_j(x),2n)+o(x^{p_j}))^{d_{ij}}) + f_i^-(x)(\prod_{j=1}^t(taylor(trans_j(x),2n-1)+o(x^{q_j}))^{d_{ij}}))\\
  \ \ \  =\sum_{i=1}^s(f_i^{+}(x)(\prod_{j=1}^t(taylor(trans_j(x),2n))^{d_{ij}}) + f_i^-(x)(\prod_{j=1}^t(taylor(trans_j(x),2n-1))^{d_{ij}})) + o(x^{dd_n})$,

  where for \ $j=1\cdots t,p_j\geq 2n$, \ $q_j\geq 2n-1$, \ $dd_n\geq \min\{\min\{p_j,q_j\}| i=1,\cdots\}\geq(2n-1)$, that is to say,\ $F(x)=\operatorname{T_{min}}(n,F)+o(x^{dd_n})$, \ $dd_n\geq 2n-1$.

  By Lemma 3.5, there exists \ $n_0\in \mathbb{N}$ such that \ $td(\operatorname{T_{min}}(n_0,F))<2n_0-1$, in other words,\ $\operatorname{T_{min}}(n_0,F)=c_{n_0} x^{d_{n_0}}+o(x^{d_{n_0}})$, where \ $d_{n_0}<2n_0-1$ and \ $c_{n_0}\neq 0$.

  Let \ $n=n_0$, then \ $F(x)=\operatorname{T_{min}}(n_0,F)+o(x^{dd_{n_0}})$,\ $dd_{n_0}\geq 2n_0-1$.

  We get that \ $F(x)= c_{n_0}x^{d_{n_0}}+o(x^{d_{n_0}})+ o(x^{dd_{n_0}})= cn_{0} x^{d_{n_0}}+ o(x^m)$, where \ $m=\min\{d_{n_0},dd_{n_0}\}=d_{n_0}$.

  Let \ $\overline{F}(x)= F(x)/x^m = c_{n_0}+o(x^m)/ x^m $, then for \ $x\rightarrow 0^{+}$, \ $F(x)\rightarrow c_{n_0}$. As \ $F(x)>0$ holds on \ $(0,T)$, so \ $\overline{F}(x)>0$ holds on \ $(0,T)$, we have \ $c_{n_0}\geq 0$. \ as \ $c_{n_0}\neq 0$, we get that \ $c_{n_0}>0$.

  So \ $\operatorname{T_{min}}(n_0,F)=c_{n_0} x^m+o(x^m)$, where \ $c_{n_0}>0$. Then for \ $x\rightarrow 0^+$,\ $H(x)=\operatorname{T_{min}}(n_0,F)/x^m = c_{n_0}+o(x^m)/x^m \rightarrow c_{n_0}>0$, thus there exists \ $ \delta\in (0,T)$, for \  $x\in (0,\delta)$, \ $H(x)= \operatorname{T_{min}}(n_0,F)/x^m>0$, so \ $\operatorname{T_{min}}(n_0,F)=H(x) x^m>0$.

  We claim that the lemma holds.
  \end{proof}
\end{lemma}

\begin{lemma}\label{lemma3.7}
Let \ $F(x)=f(x,trans_1(x),\cdots,trans_t(x))>0$, $trans_1(x),\cdots, trans_t(x)$ can be regularly expanded on \ $(0,b]$, then \ $F(x)>0$ holds on \ $[a,b]$ if and only if there exists an \ $n_0$ such that \ $\operatorname{T_{min}}(n_0,F)(x)>0$ holds on \ $[a,b]$, where \ $b>a>0$.
  \begin{proof}
   Let \ $\varepsilon$ be the minimum value of \ $F(x)$ on the close interval \ $[a,b]$, then \ $\varepsilon>0$ obviously.

  When \ $n\rightarrow\infty$, \ $taylor(trans_j(x),n)$ converges uniformly to \ $trans_j(x)$ on \ $[a,b]$ for \ $j=1,\cdots,t$, thereby \ $\operatorname{T_{min}}(n,F)(x)$ converge uniformly to \ $F(x)$ on \ $[a,b]$. So there exists \ $n_0$ such that \ $|F(x)-\operatorname{T_{min}}(n_0,F)(x)|<\frac{\varepsilon}{2}$  on \ $[a,b]$. Then we get that \ $\operatorname{T_{min}}(n_0,F)(x)>F(x)-\frac{\varepsilon}{2} > \frac{\varepsilon}{2}$.
  \end{proof}
\end{lemma}

\begin{theorem}
Let  $F(x)=f(x,trans_1(x),\cdots,trans_t(x))>0$, \ $trans_1(x),\cdots,trans_t(x)$ can be regularly expanded on \ $(0,T]$, then \ $F(x)>0$ holds on \ $(0,T]$ if and only if there is \ $n_0$ such that \ $\operatorname{T_{min}}(n_0,F)(x)>0$ holds on \ $(0,T]$.
\begin{proof}
The sufficiency holds obviously.

By Lemma 3.6, there exist \ $n_1$ and \ $\delta\in (0,T)$ such that \ $\operatorname{T_{min}}(n,F)(x)>0$ holds on \ $(0,\delta)$ for \ $n\geq n_1$.

By Lemma 3.7, there exists \ $n_2$ such that \ $\operatorname{T_{min}}(n,F)(x)>0$ holds on \ $[\frac{\delta}{2}, T]$ for \ $n\geq n_2$.

Let \ $n_0 = \max\{n_1,n_2\}$, then \ $\operatorname{T_{min}}(n,F)(x)>0$ holds on \ $(0,T]$ for \ $n\geq n_0$.
\end{proof}
\end{theorem}

\begin{corollary}
Let  $F(x)=f(x,trans_1(x),\cdots,trans_t(x))>0$, \ $trans_1(x),\cdots,trans_t(x)$ can be regularly expanded on \ $(0,T]$, then \ $F(x)<0$ holds on \ $(0,T]$ if and only if there is \ $n_0$ such that \ $\operatorname{T_{max}}(n_0,F)(x)<0$ holds on \ $(0,T]$.
\end{corollary}

\begin{lemma}
For a real differentiable function \ $F(x)$ and \ $x_0\in \mathbb{R}$, if \ $F(x_0) = 0$ and \ $F(x)$ is greater or less than $0$ in a deleted neighborhood of \ $x_0$, then \ $F'(x_0) = 0$.
\end{lemma}

 A deleted neighborhood of \ $x_0$ is a neighborhood of \ $x_0$ with the point \ $x_0$ removed.

\begin{lemma}\label{lemma3.9}
For real differentiable function \ $F(x)$, if \ $F(x)$ has no multiple real root, then \ $F(x)>0$ and \ $F(x)\geq 0$ are equivalent, \ $F(x)<0$ and \ $F(x)\leq 0$ are equivalent.
\begin{proof}
Assume that \ $F(x)\geq 0$ holds but \ $F(x)>0$ doesn't, then there exists \ $x_0$ such that \ $F(x_0)=0$ and \ $F(x)>0$ on a  deleted neighborhood of \ $x_0$ ,then \ $F'(x_0)=0$, which contradicts with the assumption.

Similarly, \ $F(x)<0$ is equivalent to \ $F(x)\leq 0$.
\end {proof}
\end{lemma}

\begin{theorem}\label{theorem3.5}
For $F(x)=f(x,trans_1(x),\cdots,trans_t(x))>0$, if \ $trans_1(x),\cdots,trans_t(x)$ can be regularly expanded on \ $(0,T]$ and \ $F(x)$ has no multiple real root, then Algorithm 2 must terminate.
\begin{proof}
If \ $F(x) > 0$ or \ $F(x)< 0$ holds, then the algorithm will terminate due to Theorem 3.4 and Corollary 3.1.

If neither \ $F(x) > 0$ nor \ $F(x) < 0$ holds, then neither \ $F(x)\geq 0$ nor \ $F(x)\leq 0$ holds by Lemma 3.9. So there exist \ $x_1$ such that \ $F(x_1) < 0$ and \ $x_2$ such that \ $F(x_2) > 0$. By Theorem 3.3, there exist \ $n_1$ such that \ $\operatorname{T_{max}}(n_1, f)(x_1)\leq 0$ and \ $n_2$ such that \ $\operatorname{T_{min}}(n_2, f)(x_2)\geq 0$, so the algorithm is bound to terminate when \ $n = \max\{n_1,n_2\}$.
\end{proof}
\end{theorem}

\section{Reachability Analysis of Linear Systems }

In this section we aim to:
\begin{enumerate}
\item decide the reachability of linear system whoes initial set contains only one point;

\item decide the reachability of linear system whoes initial set is an open semi-algebraic set.
\end{enumerate}

For the first case, we will decide the sign of the respective trigonometric-exponential polynomial directly by Successive Taylor-substitution. For the second case, we will propose a decision procedure based on openCAD (See ref[13]) and an algorithm of real root isolation derivated from Successive Taylor-substitution.

\subsection{Decision procedure for the initial set containing only one point}

From the above discussion, if  the initial set contains only one point, to decide the safety of linear system we can resort to the sign-deciding of the transcendental function of a class of  trigonometric-exponential polynomial.

As \ $e^{x}$ can not be regularly expanded on \ $(0,T)$ for \ $\forall T>0$, while \ $e^{-x}$ can, we need the following form of trigonometric-exponential polynomial.

 \ $F(x)=f(x,e^{-u_{1}x},\cdots,e^{-u_{r}x},\sin(v_{1}x),\cdots,\sin(v_{s}x),\cos(r_{1}x),\cdots,\cos(r_{t}x))$, where \ $u_i,v_i,r_i\in \mathbb{R}_{Alg}$ and \ $u_i>0$ for \ $i=1,\cdots,r$.

As \ $\sin(x)$ can not be regularly expanded on \ $(0,T)$ for \ $T\geq \pi$, \ $\cos(x)$ can not be regularly expanded on \ $(0,T)$ for \ $T\geq \frac{\pi}{2}$, we assume that \ $\max\{v_1,\cdots,v_s\}<\frac{\pi}{T}$, \ $\max\{r_1,\cdots,r_t\}<\frac{\pi}{2 T}$ with the help of Duplication Formulae $\cos(x)=1-2\sin(\frac{x}{2})^2$ and \ $\sin(x)=2\sin(\frac{x}{2})\cos(\frac{x}{2})$ to guarantee that all \ $\sin(v_{i} x)$s and \ $\cos(r_{i} x)$s can be expanded regularly on \ $(0,T)$.

\begin{algorithm}[H]
\caption{$Decide\_trigonometric\_exponential\_polynomial$}
\LinesNumbered 
\KwIn{$F(x)=f(x,e^{-u_{1}x},\cdots,e^{-u_{r}x},\sin(v_{1}x),\cdots,\sin(v_{s}x),\cos(r_{1}x),\cdots,\cos(r_{t}x))$, and a constant \ $T$ such that \ $T\times \max\{v_1,\cdots,v_s\}< \pi$ and \ $T\times \max\{r_1,\cdots,r_t\}<\frac{\pi}{2}$}
\KwOut{The sign of \ $F(x)$ on \ $I$}

$n_1\leftarrow \max\{n_0(e^{-u_{i} x}, T)|i=1,\cdots,r\}$

$n_2\leftarrow \max\{n_0(\sin(v_{i} x), T)|i=1,\cdots,s\}$

$n_3\leftarrow \max\{n_0(\cos(r_{i} x), T)|i=1,\cdots,t\}$

$n_0\leftarrow \max\{n_1, n_2, n_3\}$

$n\leftarrow [n_{0}/2]+1$, where \ $[x]$ denotes the biggest integer less than or equal to $x$

$\operatorname{T_{max}}(n,F) \leftarrow f^{+}(x, taylor(e^{-u_{1} x},2n-1), \cdots, taylor(e^{-u_{r} x},2n-1), taylor(\sin(v_{1} x),2n-1), \cdots,$

$taylor(\sin(v_{s} x), 2n-1), taylor(\cos(r_{1} x), 2n-1), \cdots, taylor(\cos(r_{t} x),2n-1)) + f^{-}(x, taylor(e^{-u_{1} x},$x $2n), \cdots, taylor(e^{-u_{r} x}, 2n), taylor(\sin(v_{1} x), 2n), \cdots, taylor(\sin(v_{s} x), 2n), taylor(\cos(r_{1}x), 2n), \cdots, taylor(\cos(r_{t} x), 2n))$

$\operatorname{T_{min}}(n,F) \leftarrow f^{+}(x, taylor(e^{-u_{1} x}, 2n), \cdots, taylor(e^{-u_{r} x}, 2n), taylor(\sin(v_{1} x), 2n), \cdots, taylor(\sin(v_{s}x),2n), taylor(\cos(r_{1} x),2n), \cdots, taylor(\cos(r_{t} x), 2n)) + f^{-}(x,taylor(e^{-u_{1} x}, 2n-1),\cdots,taylor(e^{-u_{r} x},2n-1), taylor(\sin(v_{1} x), 2n-1), \cdots, taylor(\sin(v_{s} x), 2n-1), taylor(\cos(r_{1} x), 2n-1), \cdots, taylor(\cos(r_{t}x), 2n-1))$

\If{$\operatorname{T_{min}}(n,F)\geq 0$ holds on \ $(0,T]$}{\Return{1} \tcp{$F(x)>0$ holds on $(0,T]$}}

\If {$\operatorname{T_{max}}(n,F)\leq 0$ holds on \ $(0,T]$}{\Return{-1}   \tcp{$F(x)>0$ holds on $(0,T]$}}

\eIf {neither $\operatorname{T_{max}}(n,F)\geq 0$ nor $\operatorname{T_{min}}(n, F)\leq 0$ holds on $(0,T]$}{\Return{0} \tcp{$F(x)$ has no constant sign on $(0,T]$}}{$n\leftarrow n+1$, go to 6)}

\end{algorithm}

Algorithm 3 is the application of Algorithm 2 to trigonometric-exponential polynomial, so we have the following corollary from Theorem 3.5.

\begin{corollary}
If \ $F(x)$ has no multiple real root, Algorithm 3 must terminate.
\end{corollary}

For the general trigonometric-exponential polynomial, we designed the following algorithm based on Algorithm 3 and with the help of Algorithm 1.

\begin{algorithm}[H]
\caption{$Decide\_reducible\_trigonometric\_exponential\_polynomial$}
\LinesNumbered 
\KwIn{A trigonometric-exponential polynomial \ $F(x)$  and a constant $T$}
\KwOut{The sign of \ $F(x)$ on \ $(0,T]$}

Run Algorithm 1 to decompose \ $F(x)$ to \ $F(x)=C\times F_1(x)^{n_1}\times \cdots\times F_m(x)^{n_m}$ such that \ $F_1(x),\cdots,F_m(x)$ are pairwisely co-prime and each \ $F_i(x)$ has no  multiple real root

Run Algorithm 3 to decide the signs of \ $F_1(x),\cdots,F_m(x)$ on  $(0,T]$ respectively

Get the sign of $F(x)$ on $(0,T]$\;
\end{algorithm}

The following example is to show how to decide the reachability of linear system by Algorithm 3.

\begin{example}
(Adapted from ref[11]) There are three reservoirs No.1, No.2 and No.3, which are connected by water pipes. There is an external pollution source connected with No.1 pool, which continuously diffuses pollutants into No.1 pool, and the pollutants diffused into No.2 and No.3 pools through connecting water pipes. Denote the amount of pollutants in No.1, 2 and 3 reservoirs by \ $x_1(t)$, \ $x_2(t)$ and \ $x_3(t)$ respectively, with the unit as pounds. $t$ represents time in minutes. It is assumed that the pollutants in each pool are evenly mixed, and the external pollution source diffuses pollutants to the No.1 pool at a speed of \ $0.01$ pounds per minute. It is also assumed that the diffusion equation of pollutants in the three pools is as follows:

$x_{1}’(t)=0.001 x_{3}(t)-0.001 x_{1}(t)+0.01$,

$x_{2}'(t)=0.001 x_{1}(t)-0.001 x_{2}(t)$,

$x_{3}'(t)=0.001 x_{2}(t)-0.001 x_{3}(t) )$,

$x_1 (0)=x_2 (0)=x_3 (0)=0$.

The unsafe set \ $Y = \{(y_1, y_2, y_3)^T | y_2 -y_3 + 6 < 0\}$.

The eigenvalues of the matrix are $0, \frac{3}{2000} + \frac{\sqrt3}{2000}I,-\frac{3}{2000}-\frac{\sqrt3}{2000}I $, the solution of the system is:

$x_1(t) = \frac{10\sqrt{3}}{9} e^{-\frac{3t}{2000}} \sin(\frac{\sqrt{3} t}{2000}) - \frac{10}{9} e^{-\frac{3t}{2000}} \cos(\frac{\sqrt{3} t}{2000}) + \frac{t}{300} + \frac{10}{3}$;

$x_2(t) = -\frac{20\sqrt{3}}{9} e^{-\frac{3t}{2000}} \sin(\frac{\sqrt{3}t}{2000}) +\frac{t}{300}$;

$x_3 (t)=\frac{10\sqrt{3}}{9} e^{-\frac{3t}{2000}} \sin(\frac{\sqrt{3} t}{2000})-\frac{10}{9} e^{-\frac{3t}{2000}} \cos(\frac{\sqrt{3} t}{2000})+\frac{t}{300}-\frac{10}{3}$.

Thus, the problem is transformed to decide whether the inequality
$F(t)=y_{2}-y_{3}+6 = \\ -\frac{10\sqrt{3}}{3} e^{-\frac{3t}{2000}} \sin(\frac{\sqrt{3} t}{2000})-\frac{10}{3} e^{-\frac{3t}{2000}} \cos(\frac{\sqrt{3} t}{2000})+\frac{28}{3}>0$ holds on \ $(0, T)$. That the inequality holds implies the system is safe, otherwise the system is unsafe.

(1) Assume that \ $T=1000$, the thresholds \ $n_0(e^{-\frac{3t}{2000}}, 1000)=2$, $n_0(\sin(\frac{\sqrt{3} t}{2000}), 1000)=1$, \ $n_0(\cos(\frac{\sqrt{3} t}{2000}), 1000)=1$, where $xprove(\operatorname{T_{min}}(F, 2)>0, [t<1000])= true$, where \ $xprove$ is an inequality-proving tool in BOTEMMA-package, that is to say, \ $F(t)>0$ holds on \ $(0,1000)$ and then the system is safe for \ $t<1000$.

We have implemented the algorithm on Maple16, which have been run on a 64-bit Hp computer with an Intel(R) Core(TM)2 Quad CPU @ 2.66GHz 2.67GHz and 3GB of RAM.
The time consumption is \ $0.14$ seconds (The algorithms' running environment is the same and time consumption in seconds is denoted by TC in the sequel).

(2) Assume that \ $T=2000$, the thresholds \ $n_0(e^{-\frac{3t}{2000}}, 2000)=4$, $n_0(\sin(\frac{\sqrt{3} t}{2000}), 2000)=1$, but \ $\cos(\frac{\sqrt{3} t}{2000})$ cannot be expanded regularly on \ $(0,2000)$.

By Duplication Formula \ $\cos(\frac{\sqrt{3} t}{2000}=1-(\sin(\frac{\sqrt{3} t}{4000}))^2$, we have that \ $F(t)=y_{2}-y_{3}+6 = -\frac{10 \sqrt{3}}{3} e^{-\frac{3 t}{2000}} \sin(\frac{\sqrt{3} t}{2000}) + \frac{10}{3} e^{-\frac{3 t}{2000}} (\sin(\frac{\sqrt{3} t}{4000}))^2-\frac{10}{3} e^{-\frac{3 t}{2000}}+\frac{28}{3}>0$ and \ $n_0(\sin(\frac{\sqrt{3} t}{4000}), 2000)=1$.

Then as $xprove(\operatorname{T_{min}}(F, 4)>0, [t<2000])= true$, we get that, \ $F(t)>0$ holds on \ $(0,2000)$ and then the system is safe for \ $t<2000$.

\ $TC=0.013$.

\end{example}

\subsection{Decision procedure for the open semi-algebraic initial set}

When the initial set is a semi-algebraic set, Algorithm 3 is an optional heuristic scheme.

\begin{example}

Let us continue Example 3 with the initial set revised as \ $T = \{(t_1, t_2, t_3)^T | (t_1 - 1)^2 + (t_2- 1)^2 + (t_3 -1)^2 < 1\}$, and the other contents remaining the same.

In this case, we need to show whether the following trigonometric-exponential polynomial inequality with parameters \ $t_1, t_2, t_3$ holds on the condition that \ $(t_1 - 1)^2 + (t_2- 1)^2 + (t_3 -1)^2 < 1$,

\noindent\ $F(t)=-\frac{10}{3} \sqrt{3} a b+ \frac{2}{3} \sqrt{3} t_1 a b+ t_2 a c-t_3 a c- \frac{\sqrt{3}}{3} t_3 a b -\frac{10}{3} a c- \frac{\sqrt{3}}{3} t_2 a b +\frac{28}{3}>0$,
where
\ $a=e^{-\frac{3}{2000} t}$, \ $b=\sin(\frac{\sqrt{3}}{2000} t)$, \ $c = \cos(\frac{\sqrt{3}}{2000} t)$.

By BOTTEMA-package, we have \ $xprove(T\_{min}(f,2)>0,[(t_1-1)^2+(t_2-1)^2+(t_3-1)^2<1,t<1000])=true$ and then we get that the system is safe for \ $t<1000$.

\ $TC=14.765$.
\end{example}

 For more general cases, Cylindrical Algebraic Decomposition (CAD) is needed.

The basic idea of CAD is as follows: given a set $S$ of polynomials in \ $\mathbb{R}[x]$, CAD can be used to partition \ $\mathbb{R}^n$ into connected semi-algebraic sets, called cells, such that each polynomial in $S$ keeps constant sign (either $+$, $-$ or $0$) on each cell. When constraints are open sets, GCAD (ref[23]) or openCAD (ref[13]) is enough, which partitions the space $\mathbb{R}^n$ into a set of open cells instead of cells (i.e., takes sample points from open cells only), such that on each of which every polynomial in S keeps constant non-zero sign (either $+$ or $-$).

The CAD procedure needs the real root isolation of TEP as its basic algorithm. We give a real root isolation algorithm based on the sign-deciding procedure Algorithm 3 or Algorithm 4. The basic idea is as follows:

If \ $F(x)>0$ or \ $F(x)<0$ holds on \ $(a, b)$, we get that \ $F(x)$ has no real root. Otherwise, if \ $F(x)$ is monotonous on \ $(a, b)$, then \ $F(x)$ has one and only one root on \ $(a, b)$, else dividing the interval \ $(a, b)$ repeatedly by dichotomy until on each interval there is either a unique real root or no real root.

The most effective way to decide the monotonicity of \ $F(x)$ is to determine weather \ $F'(x) >0$ or \ $F'(x) <0$ holds. What should be pointed out is that the derivative of the trigonometric-exponential polynomial is also a trigonometric-exponential polynomial.

We need modify Algorithm 3 slightly as follow.

\begin{algorithm}[H]
\caption{$Decide\_trigonometric\_exponential\_polynomial(DTEP)$}
\LinesNumbered 
\KwIn{$F(x)=f(x,e^{-u_{1}x},\cdots,e^{-u_{r}x},\sin(v_{1}x),\cdots,\sin(v_{s}x),\cos(r_{1}x),\cdots,\cos(r_{t}x))$, and an interval \ $(a, b]$, where \ $b>a\geq 0$, \ $b\times \max\{v_1, \cdots, v_s\}< \pi $ and \ $b\times \max\{r_1, \cdots , r_t\}< \frac{\pi}{2}$}
\KwOut{The sign of \ $F(x)$ on \ $(a, b]$}

$n_1\leftarrow \max\{n_0(e^{-u_{i} x}, b)|i=1,\cdots,r\}$

$n_2\leftarrow \max\{n_0(\sin(v_{i} x), b)|i=1,\cdots,s\}$

$n_3\leftarrow \max\{n_0(\cos(r_{i} x), b)|i=1,\cdots,t\}$

$n_0\leftarrow \max\{n_1, n_2, n_3\}$

$n \gets [\frac{n_{0}}{2}]+1$

\If{$\operatorname{T_{min}}(n, F)\geq 0$ holds on $(a, b]$}{\Return{1} \tcp{$F(x)>0$ holds on $(a, b]$}}

\If{$\operatorname{T_{max}}(n, F)\leq 0$ holds on $(a, b]$}{\Return{-1}  \tcp{$F(x)<0$ holds on $(a, b]$}}

\eIf{neither $\operatorname{T_{max}}(n, F)\geq 0$ nor $\operatorname{T_{min}}(n, F)\leq 0$ holds on $(a, b]$}{\Return{0}  \tcp{$F(x)<0$ holds on $(a, b]$}}{$n \gets n+1$, goto 6)}

\end{algorithm}

The computations of $\operatorname{T_{min}}(n, F)$ and $\operatorname{T_{max}}(n, F)$ are the same as Algorithm 3.

Based on Algorithm 5, we present an algorithm for the real roots isolation of trigonometric- exponential polynomial.

\begin{algorithm}[H]
\caption{$Isolation\_trigonometric\_exponential\_polynomial$(ITEP)}
\LinesNumbered 
\KwIn{A trigonometric-exponential polynomial \ $F(x)$ and an interval \ $(a, b)$, where\ $0 \leq a<b$ and \ $a$, $b\in \mathbb{Q}$}
\KwOut{$L=\{(a_n, b_n)\}$, where \ $\{(a_i, b_i)|i=1, \cdots , n\}$ are pairwise disjoint, \ $F(x)$ has one real root on each \ $(a_i, b_i)$, and \ $L$ contains all the real roots of \ $F(x)$ on \ $(a, b)$}

$sgn \gets DTEP(F, (a, b))$

\If {$sgn=1$ or \ $sgn=-1$}{\Return{$\Phi$} }

\eIf {$sgn=0$}{$dF \gets diff(F, x)$

$dsgn \gets DTEP(dF, (a, b))$

\If {$dsgn=1$ or \ $dsgn=-1$}{\Return{$\{(a, b)\}$}}

}{\Return{$ITEP(F, (a,\frac{a+b}{2})) \bigvee ITEP(F, (\frac{a+b}{2}, b))$}}

\end{algorithm}

Obviously, the endpoints of all intervals appearing in Algorithm 6 are rational. By Lemma 2.1, for $\operatorname{cont}(F(x))=1$ and \ $0 \neq x_0\in \mathbb{Q}$, \ $F(x_0)\neq 0$, so whether the intervals are open or close is not cared by the Algorithm 6.

To describe the process of Algorithm 6, we propose the following Example 5.

\begin{example}

Let us continue Example 2 to isolate the real roots of

$f(x)=-\sqrt3-24\sqrt3 e^{-x}-4\sin(\frac{\sqrt3}{2}x) e^{-\frac{3}{2} x}-12\sqrt3 e^{-\frac{5}{2}x}+108\sqrt3 e^{-2x}-8e^{-3x} \sin(\frac{\sqrt3}{2}x)\cos(\frac{\sqrt3}{2}x)+36e^{-\frac{5}{2}x}\sin(\frac{\sqrt3}{2} x)$ on $(0,3)$.

1)\ As $\cos(\frac{\sqrt3}{2}x)$ can not be regularly expanded on $(0,3)$, by Duplication Formula

$-\sqrt3 - 24 \sqrt3 e^{-x} - 4\sin(\frac{\sqrt3}{2}x) e^{-\frac{3}{2}x}-12\sqrt3 e^{-\frac{5}{2} x}+108\sqrt3 e^{-2x} - 8\sin(\frac{\sqrt3}{2}x) e^{-3x}+16 sin(\frac{\sqrt3}{2}x) \sin(\frac{\sqrt3}{4} x)^2 \ e^{-3x}+36 e^{-\frac{5}{2}x} \sin(\frac{\sqrt3}{2} x)$

2) The derivative of \ $f(x)$ is

$df(x) = 24 \sqrt3 e^{-x} - 2 \sqrt3 \cos(\frac{\sqrt3}{2} x) e^{-\frac{3}{2}x}+ 6 \sin(\frac{\sqrt3}{2}x) e^{-\frac{3}{2}x} + 30 \sqrt3 e^{-\frac{5}{2}x} - 216\sqrt3 e^{-2x}+24 \sin(\frac{\sqrt3}{2}x) e^{-3x}-4 \sqrt3 \cos(\frac{\sqrt3}{2}x)e^{-3x}-48\sin(\frac{\sqrt3}{2}x) sin(\frac{\sqrt3}{4}x)^2 e^{-3x} + 8\sqrt3 \cos(\frac{\sqrt3}{2}x) \sin(\frac{\sqrt3}{4}x)^2 e^{-3x}+8\sqrt3 \sin(\frac{\sqrt3}{2}x) \sin(\frac{\sqrt3}{4}x) \cos(\frac{\sqrt3}{4} x) e^{-3x}-90 e^{-\frac{5}{2}x} sin(\frac{\sqrt3}{2}x)+18 e^{-\frac{5}{2}x} \sqrt3 \cos(\frac{\sqrt3}{2} x)$

3)\ Run Algorithm 5, we have that \ $DTEP(f,(0,3)) = 0$ and \ $DTEP(df,(0,3)) = 0$.

Dichotomy yields that

\ $ DTEP(f,(0, \frac{3}{2}))= 0$ and \ $DTEP(df, (0, \frac{3}{2})) =-1$ which implies that $f(x)$ has one and only one real root on \ $(0, \frac{3}{2})$.

\ $ DTEP(f,(\frac{3}{2}, 3))= -1$, which means that \ $f(x)<0$ and \ $f(x)$ has no real root on \ $(\frac{3}{2}, 3)$.

So, we conclude that $f(x)$ has one and only one real roots on \ $(0, \frac{3}{2})$.

The total time consumption is \ $34.67s$.
\end{example}

To show how to decide the reachability of a linear system, we present Example 6.

\begin{example}
Consider the following linear system
\begin{center}
$\xi^{\prime}={\begin{bmatrix}
1 & -1 & 1 \\
1 & -1 & 0 \\
0 & 1 & 0
\end{bmatrix}} \xi+{\begin{bmatrix}
1 \\
1 \\
1
\end{bmatrix}}$
\end{center}

The initial set \ $X=\{(x_1, x_2, x_3)^T|x_1^2+x_2^2+2x_3^2<1\}$, the unsafe set \ $Y=\{(y_1, y_2, y_3)^T|y_2-y_1+5<0\}$, and we assume that \ $t \in (0,3)$.

The eigenvalues of the matrix are \ ${1,-\frac{1}{2}-\frac{\sqrt3}{2}I,-\frac{1}{2}+\frac{\sqrt3}{2}I}$ respectively, and the solution of the system is

$\xi_1 = ((\frac{2}{3} e^t+\frac{\sqrt3}{3} e^{-\frac{1}{2} t} \sin(\frac{\sqrt3}{2} t)+\frac{1}{3} e^{-\frac{1}{2} t} \cos(\frac{\sqrt3}{2} t) ) x_1-\frac{2\sqrt3}{3} e^{-\frac{1}{2} t} \sin(\frac{\sqrt3}{2} t) x_2+(-\frac{2}{3} e^{-\frac{1}{2} t} \cos(\frac{\sqrt3}{2} t)+\frac{2}{3} e^t ) x_3+\frac{4}{3} e^t+\frac{2}{3} e^{-\frac{1}{2} t} \cos(\frac{\sqrt3}{2} t)-2$,

$\xi_2=(\frac{1}{3} e^t+\frac{\sqrt3}{3} e^{-\frac{1}{2} t} \sin(\frac{\sqrt3}{2} t)-\frac{1}{3} e^{-\frac{1}{2} t} \cos(\frac{\sqrt3}{2} t) ) x_1+(-\frac{\sqrt3}{3} e^{-\frac{1}{2} t} \sin(\frac{\sqrt3}{2} t)+\frac{1}{3} e^{-\frac{1}{2} t} \cos(\frac{\sqrt3}{2} t) ) x_2+(\frac{\sqrt3}{3} e^{-\frac{1}{2} t} \sin(\frac{\sqrt3}{2} t)-\frac{1}{3} e^{-\frac{1}{2} t} \cos(\frac{\sqrt3}{2} t)+\frac{1}{3} e^t ) x_3+\frac{\sqrt3}{3} e^{-\frac{1}{2} t} \sin(\frac{\sqrt3}{2} t)+\frac{1}{3} e^{-\frac{1}{2} t} \cos(\frac{\sqrt3}{2} t)+\frac{2}{3} e^t-1$,

$\xi_3=(\frac{1}{3} e^t-\frac{\sqrt3}{3} e^{-\frac{1}{2} t} \sin(\frac{\sqrt3}{2} t)-\frac{1}{3} e^{-\frac{1}{2} t} \cos(\frac{\sqrt3}{2} t) ) x_1+\frac{2\sqrt3}{3} e^{-\frac{1}{2} t} \sin(\frac{\sqrt3}{2} t) x_2+(\frac{2}{3} e^{-\frac{1}{2} t} \cos(\frac{\sqrt3}{2} t)+\frac{1}{3} e^t ) x_3+\frac{2}{3} e^t$.

Thus, the reachability problem becomes \ $\Gamma = $

$\exists t \in (0,3), \exists x_1, \exists x_2,\exists x_3, \phi(t,x_1,x_2,x_3) =(-\frac{2}{3} e^{-\frac{1}{2} t} \cos(\frac{\sqrt3}{2} t) - \frac{1}{3} e^{t} ) x_1+(\frac{\sqrt3}{3} e^{-\frac{1}{2} t} \sin(\frac{\sqrt3}{2} t) + e^{-\frac{1}{2} t} \cos(\frac{\sqrt3}{2} t) ) x_2 + (-\frac{\sqrt3}{3} e^{-\frac{1}{2} t} \sin(\frac{\sqrt3}{2} t) + \frac{1}{3} e^{-\frac{1}{2} t} \cos(\frac{\sqrt3}{2} t)-\frac{1}{3} e^t ) x_3 + \frac{\sqrt3}{3} e^{-\frac{1}{2} t} \sin(\frac{\sqrt3}{2} t) - \frac{1}{3} e^{-\frac{1}{2} t} \cos(\frac{\sqrt3}{2} t)-\frac{2}{3} e^t+6 < 0$.

Denote \ $ a=e^{-\frac{1}{2}t}$, $b=\sin(\frac{\sqrt3}{2}t)$, $c=\cos(\frac{\sqrt3}{2}t)$, $d=e^{t}$, using Brown’s projection operator to eliminate \ $x_1$,\ $x_2$,\ $x_3$ successively , we have

$q_0 (t,x_1,x_2,x_3)=(x_1^2+x_2^2+2x_3^2-1)((-\frac{2}{3} ac-\frac{1}{3} d) x_1+(\frac{\sqrt3}{3} ab+ac) x_2+(-\frac{\sqrt3}{3} ab+\frac{1}{3} ac-\frac{1}{3} d) x_3+\frac{\sqrt3}{3} ab-\frac{1}{3} ac-\frac{2}{3} d+6)$,

$q_1 (t,x_1,x_2)=q_{11} \times q_{12} \times q_{13}$,

$q_2 (t,x_1)=q_{21} \times q_{22}\times q_{23}\times q_{24} \times q_{25}\times q_{26}$,

$q_3 (t)=a \times q_{31} \times q_{32} \times q_{33} \times q_{34} \times q_{35} \times q_{36} \times q_{37} \times q_{38} \times q_{39}$,\\where

$q_{11}=x_1^2+x_2^2-1$,

$q_{12}=-\frac{\sqrt3}{3} ac+\frac{\sqrt3}{3} d+ab$,

$q_{13}=216-8a^2 c^2 x_1 x_2+2acdx_1^2-\frac{2}{3} acdx_2^2+\frac{20}{3} acdx_1-8acdx_2+24\sqrt3 abx_2-\frac{2\sqrt3}{3} a^2 bc-\frac{10\sqrt3}{3} abd-\frac{8\sqrt3}{3} a^2 bcx_1 x_2-\frac{4\sqrt3}{3} abdx_1 x_2+24\sqrt3ab-48d-\frac{2\sqrt3}{3}a^2 bc x_1^2+a^2 b^2 x_1^2+\frac{7}{3}d^2+3a^2 b^2 x_2^2+4a^2 b^2 x_2+3a^2 c^2 x_1^2+\frac{19}{3} a^2 c^2 x_2^2+ \frac{8}{3} a^2 c^2 x_1-4a^2 c^2 x_2+\frac{10}{3} acd-48acx_1+72acx_2-4acdx_1x_2+\frac{10\sqrt3}{3} a^2 bcx_2^2-\frac{8\sqrt3}{3}a^2bcx_1+\frac{8\sqrt3}{3}a^2bcx_2+\frac{2\sqrt3}{3} abdx_1^2+\frac{2\sqrt3}{3} abdx_2^2-\frac{4\sqrt3}{3} abdx_1-\frac{8\sqrt3}{3} abdx_2+\frac{1}{3}a^2 c^2+d^2x_1^2+ \frac{1}{3}d^2x_2^2+ \frac{8}{3}d^2x_1+a^2 b^2-24dx_1-24ac$,

$q_{21}=x_1-1$,

$q_{22}=x_1+1$,

$q_{23}=-\frac{\sqrt3}{3} ac+\frac{\sqrt3}{3} d+ab$,

$q_{24}=\frac{10\sqrt3}{9} a^2 bc+\frac{2\sqrt3}{9} abd+a^2 b^2+\frac{19}{9} a^2 c^2-\frac{2}{9} acd+\frac{1}{9} d^2$,

$q_{25}=2\sqrt3 a^2 bcx_1^2-\frac{4\sqrt3}{3} a^2 bcx_1+a^2 b^2 x_1^2+\frac{13}{3} a^2 c^2 x_1^2-\frac{8\sqrt3}{3} a^2 bc-\frac{2\sqrt3}{3} abdx_1+\frac{4}{3} a^2 c^2 x_1+\frac{4}{3} acdx_1^2-\frac{4\sqrt3}{3} abd-\frac{8}{3} a^2 c^2+\frac{10}{3} acdx_1+\frac{1}{3} d^2 x_1^2+12\sqrt3 ab+\frac{4}{3} acd-24acx_1+\frac{4}{3} d^2 x_1-12ac+\frac{4}{3} d^2-12dx_1-24d+108$,

$q_{26}=\frac{10\sqrt3}{9} a^2 bcx_1^2-\frac{8\sqrt3}{9} a^2 bcx_1+\frac{2\sqrt3}{9} abdx_1^2+a^2 b^2 x_1^2+3a^2 c^2 x_1^2-\frac{14\sqrt3}{9} a^2 bc-\frac{4\sqrt3}{9} abdx_1+\frac{8}{9} a^2 c^2 x_1+\frac{2}{3} acdx_1^2-\frac{10\sqrt3}{9} abd-\frac{1}{3} a^2 b^2-\frac{17}{9} a^2 c^2+\frac{20}{9} acdx_1+\frac{1}{3} d^2 x_1^2+8\sqrt3 ab+\frac{10}{9} acd-16acx_1+\frac{8}{9} d^2 x_1-8ac+\frac{7}{9} d^2-8dx_1-16d+72$,

$q_{31}=\sqrt3 c+b$,

$q_{32}=-\frac{\sqrt3}{3} ac+\frac{\sqrt3}{3} d+ab$,

$q_{33}=\frac{\sqrt3}{3} ac-\frac{\sqrt3}{3} d+ab+6\sqrt3$,

$q_{34}=-\sqrt3 ac-\sqrt3 d+ab+6\sqrt3$,

$q_{35}=2\sqrt3 a^2 bc+a^2 b^2+\frac{13}{3} a^2 c^2+\frac{4}{3} acd+\frac{1}{3} d^2$,

$q_{36}=\frac{10\sqrt3}{9} a^2 bc+\frac{2\sqrt3}{9} abd+a^2 b^2+3 a^2 c^2+\frac{2}{3} acd+\frac{1}{3} d^2$,

$q_{37}=\frac{10\sqrt3}{9} a^2 bc+\frac{2\sqrt3}{9} abd+a^2 b^2+\frac{19}{9} a^2 c^2-\frac{2}{9} acd+\frac{1}{9} d^2$,

$q_{38}=\frac{\sqrt3}{2} a^2 c^2+a^2 bc+\frac{3\sqrt3}{2} ac-\frac{\sqrt3}{8} d^2+\frac{1}{2} abd+3\sqrt3 d-\frac{9}{2} ab-\frac{27\sqrt3}{2}$,

$q_{39}=\frac{14\sqrt3}{3} a^2 bc+\frac{10\sqrt3}{3} abd+a^2 b^2+\frac{25}{3} a^2 c^2-24\sqrt3 ab-\frac{2}{3} acd+24ac-\frac{5}{3} d^2+48d-216$.

Isolate all real roots of \ $q_3 (t)=0$ on \ $(0,3)$, we get that \ $q_3 (t)$ has 6 roots which are located in \ $(\frac{7113}{4096},\frac{1779}{1024})$, \ $(\frac{117}{64},\frac{3747}{2048})$, \ $(\frac{7497}{4096},\frac{1875}{1024})$, \ $(\frac{309}{128},\frac{39}{16})$, \ $(\frac{47403}{16384},\frac{94809}{32768})$,\ $(\frac{23703}{8192},\frac{47409}{16384})$ respectively.

Lift the real root isolation in the order \ $t,x_1,x_2,x_3$ successively using the openCAD lifting procedure, finally we obtain \ $95$ sample points, and \ $(\frac{29}{16},\frac{3}{4},\frac{1}{512},\frac{55}{128}))$ satisfies \ $y_2-y_1+5<0 $, which implies that the safety property is not satisfied with the counter example starting from \ $(\frac{3}{4},\frac{1}{512},\frac{55}{128})\in X $,
and ending at time \ $t = \frac{29}{16}$.
\end{example}

\section{Conclusion}
We propose a decision procedure of reachability for a class of linear system \ $\xi'=A\xi+u$, with restrictions that the matrix \ $A$ has arbitrary algebraic eigenvalues, the input \ $u$ is a vector of trigonometric-exponential polynomials. If the initial set of the linear system contains only one point, the reachability problem under consideration is resorted to the decidability of the sign of trigonometric-exponential polynomial and achieved by being reduced to verification of univariate polynomial inequalities through Taylor Expansion of the related exponential functions and trigonometric functions. If the initial set is open semi-algebraic, a decision procedure is proposed  based on openCAD and an algorithm of real roots isolation derivated from the sign-deciding procedure. The experimental results indicate the efficiency of our approach. Furthermore, the procedure is complete under the assumption of Schanuel’s Conjecture.

In addition, some algorithms presented in this paper, such as the factorization without multiple roots and Successive Taylor Substitution, are much useful for similar problems.

\bibliographystyle{elsarticle-harv}

\end{document}